\documentclass[12pt]{article}
\usepackage{a4wide}
\usepackage{slashed}
\usepackage{epsfig,graphicx}
\usepackage{amsmath}
\usepackage{amssymb}
\usepackage{array}
\usepackage{color}
\newcommand{\beq}{\begin{equation}}
\newcommand{\eeq}{\end{equation}}
\newcommand{\bea}{\begin{eqnarray}}
\newcommand{\eea}{\end{eqnarray}}
\newcommand{\beqa}{\begin{eqnarray}}
\newcommand{\eeqa}{\end{eqnarray}}
\newcommand{\cL}{{\cal L}} 
\newcommand{\Q}{{\cal Q}}

\newcommand{\nn}{\nonumber \\} 
\newcommand{\ba}{\begin{array}} 
\newcommand{\ea}{\end{array}}
\newcommand{\ol}{\overline} 
\newcommand{\dg}{\dagger} 
\newcommand{\no}{\nonumber}

\newcommand{\OMIT}[1]{{}}

\newcommand\spur{\raise.15ex\hbox{/}\kern-.57em }

\newcommand{\lsim}{
\mathrel{\hbox{\rlap{\hbox{\lower4pt\hbox{$\sim$}}}\hbox{$<$}}}}
\newcommand{\gsim}{
\mathrel{\hbox{\rlap{\hbox{\lower4pt\hbox{$\sim$}}}\hbox{$>$}}}}

\begin{document}

\begin{flushright}
LA-UR-07-4532
\end{flushright}
\vspace{1.0 true cm}
\begin{center}
{\Large {\bf
$\pi/K   \to e \bar{\nu}_e$  branching ratios  
\\
\vspace{0.5cm}
to $O(e^2 p^4)$ in Chiral Perturbation Theory}}\\
\vspace{.8 true cm}

{\large Vincenzo Cirigliano${}^{a}$ and 
Ignasi Rosell ${}^b$ } \\
\vspace{0.5 true cm}
${}^a$ 
{\sl  Theoretical Division, Los Alamos National Laboratory, Los Alamos, NM 87545, USA} \\
\vspace{0.2 true cm}
${}^b${\sl Departamento de Ciencias F\'isicas, Matem\'aticas y de la Computaci\'on,
Universidad CEU Cardenal Herrera, San Bartolom\'e 55, E-46115 Alfara del
Patriarca, Val\`encia,  Spain}\\
\vspace{0.2 true cm}
\end{center}
\vspace{0.5cm}

\begin{abstract}
We calculate  the ratios 
$R_{e/\mu}^{(P)}  \equiv 
 \Gamma( P \to e \bar{\nu}_e [\gamma] )/ \Gamma( P \to \mu \bar{\nu}_\mu [\gamma] )$    ($P=\pi,K$)  
in  Chiral Perturbation Theory  to order $e^2 p^4$.    
We complement the one- and two-loop effective theory results 
with a matching calculation  of the local counterterm, performed within  the large-$N_C$ expansion. 
We find  $R_{e/\mu}^{(\pi)} = ( 1.2352 \pm 0.0001 ) \times 10^{-4}$  
and  $R_{e/\mu}^{(K)} = ( 2.477 \pm 0.001 ) \times 10^{-5}$, with uncertainty induced by 
the matching procedure and chiral power counting.   
Given the sensitivity of upcoming new measurements,  
our results provide a clean baseline to detect or constrain effects from
weak-scale new physics  in these rare decays.  
As a by-product, we also update the theoretical analysis of the individual 
$\pi (K) \to \ell \bar{\nu}_\ell$ modes. 
\end{abstract}

\newpage

\section{Introduction}

The ratio $R_{e/\mu}^{(P)} \equiv  \Gamma( P \to e \bar{\nu}_e [\gamma] )/ 
\Gamma( P \to \mu \bar{\nu}_\mu [\gamma] )$    ($P=\pi,K$) 
of leptonic decay rates of light pseudoscalar mesons
is  helicity-suppressed in the Standard Model (SM), 
due to the $V-A$ structure of charged current couplings. 
It is therefore a  sensitive probe of  all  SM extensions 
that induce pseudoscalar currents and  
non-universal corrections to the lepton couplings~\cite{Bryman:1993gm}. 
Recently,  attention to these process has been payed in the context of the Minimal 
Supersymmetric Standard Model, 
with~\cite{Masiero:2005wr}  and 
without~\cite{Ramsey-Musolf:2007yb} lepton-flavor-violating effects. 
In general, effects from weak-scale new physics  
are expected in the range  $(\Delta R_{e/\mu})/R_{e/\mu} \sim  10^{-4} - 10^{-2}$  
and there is a realistic chance to detect or constrain them 
because of  the following circumstances. 
(i) First, ongoing experimental searches plan to reach a fractional uncertainty of 
 $(\Delta R^{(\pi)}_{e/\mu})/R^{(\pi)}_{e/\mu} \lsim  5 \times  10^{-4}$~\cite{exp-new}
and   $(\Delta R^{(K)}_{e/\mu})/R^{(K)}_{e/\mu} \lsim 3 \times 10^{-3}$~\cite{exp-new-K}, 
which represent respectively a factor of  $5$  and $10$ improvement
over current errors~\cite{PDG}. 
(ii) At the same time, the SM theoretical uncertainty  can be pushed below this level,  since to a first approximation the strong interaction dynamics cancels out in the ratio $R_{e/\mu}$ and hadronic structure dependence  
appears only through electroweak corrections.   Indeed, the most recent  
theoretical predictions read  $R^{(\pi)}_{e/\mu} = (1.2352   \pm  0.0005) \times 10^{-4}$~\cite{MS93}, 
$R^{(\pi)}_{e/\mu} = (1.2354   \pm  0.0002) \times 10^{-4}$~\cite{Fink96}, and 
$R^{(K)}_{e/\mu} = (2.472   \pm  0.001) \times 10^{-5}$~\cite{Fink96}.     
In Ref.~\cite{MS93}  a general parameterization of the hadronic effects is given, 
with an estimate of the leading model-independent contributions  
based on current algebra~\cite{terentev}. 
The dominant  hadronic  uncertainty is  roughly estimated  via  dimensional analysis. 
In Ref.~\cite{Fink96}, on the other hand,  the hadronic component is calculated by modeling 
the low- and intermediate-momentum region of the loops involving virtual photons.  
 
The primary goal of this investigation is to   
improve the current  status of the hadronic structure dependent effects. 
To this end,  we have  analyzed 
$R_{e/\mu}$   within  Chiral Perturbation Theory (ChPT)~\cite{chpt}, 
the low-energy effective field theory (EFT) of QCD.  The key feature of this  framework is that it 
provides  a controlled expansion of the amplitudes  in terms of the masses of pseudoscalar mesons and charged leptons  
($p\sim m_{\pi,K, \ell}/\Lambda_\chi$, with $\Lambda_\chi \sim 4 \pi F_\pi \sim 1.2 \,  {\rm GeV}$),  and the electromagnetic  coupling ($e$).   
Electromagnetic corrections to (semi)-leptonic 
decays of   $K$ and $\pi$ 
have been worked out to  $O(e^2 p^2)$~\cite{Knecht:1999ag,Semileptonic}, but 
had never been pushed  to  $O(e^2 p^4)$, as  required for $R_{e/\mu}$  
in order to match the experimental accuracy.  
In this work  we report full details of our  analysis of $R_{e/\mu}$ to $O(e^2 p^4)$,  
while a  summary of the results is  presented  elsewhere~\cite{short}.  
To the order we work in  ChPT,  
$R_{e/\mu}$ features 
both model independent double chiral logarithms (previously neglected) 
and an a priori unknown low-energy  coupling (LEC).  
By including  the finite loop effects and estimating the LEC via a matching calculation
in large-$N_C$ QCD,  we thus provide the first complete  result  of $R_{e/\mu}$ to $O(e^2 p^4)$ 
in  the EFT power counting.  
Most importantly, the matching calculation allows us to  further reduce the theoretical uncertainty 
and put it on more solid ground.

Our presentation is organized as follows.  In Section~\ref{sect:overview}  we introduce the basic 
definitions and outline the strategy to calculate $R_{e/\mu}$ to $O(e^2 p^4)$. 
In Section~\ref{sect:SM} we shortly review the basic ChPT formalism and the 
needed effective lagrangians. 
The loop calculation is described in Section~\ref{sect:analysis} and in 
Appendix~\ref{sect:technicalities}, and the results are reported in 
Section~\ref{sect:results}.   We  then report the matching calculation of the effective 
coupling in Section~\ref{sect:matching}, with technical details in 
Appendix~\ref{sect:matchdetails}. 
We present the contribution from real photon emission in Section~\ref{sect:real}, 
while  in Section~\ref{sect:pheno} we give our final analytical and numerical results 
for $R_{e/\mu}^{(\pi,K)}$  and discuss them. 
Section~\ref{sect:update}  is devoted to updating the theoretical expression for the 
individual  $\pi (K)  \to  \ell \bar{\nu}_\ell$ rates. 
Finally,  Section~\ref{sect:conclusions} contains our concluding remarks. 
Since we are reporting here the first ChPT calculation to order $e^2 p^4$, we give several details and 
intermediate steps of our analysis, both throughout the text and in the Appendixes.

\section{$R_{e/\mu}^{(\pi,K)}$ in ChPT: overview} 
\label{sect:overview}

To avoid excessive notational clutter,  throughout this paper we illustrate the main 
arguments in the case of  $\pi \to \ell \nu$  decays and subsequently report 
any significant changes that occur for  $K$ decays.  We consider the ratio 
\beq
R_{e/\mu}^{(\pi)}  = \frac{\Gamma \left(  \pi^+ \to e^+ \nu_e (\gamma) \right)}{\Gamma \left(  \pi^+ \to \mu^+ \nu_\mu (\gamma) \right)}
\eeq
to order $e^2 p^4$ in Chiral Perturbation Theory (ChPT).   
Within ChPT the invariant amplitudes~\footnote{ 
Intermediate steps in our analysis depend on the definition of the invariant amplitude
$T_{\ell}$  (${\ell} = \mu, e$), 
for which we use
%
$ _{\rm out} \langle    \ell^+ (p_\ell)  \nu_\ell (p_\nu)  |  \pi^+ (p) \rangle_{\rm in} =  (2 \pi)^4 \delta^{(4)} 
\left( p - p_{\ell} - p_\nu \right)  \ i \,  T_\ell $.
%
}
can be expanded in powers of the external masses and 
momenta  (of both pseudoscalar mesons and leptons) and powers of the electromagnetic coupling.  
To leading order in  the chiral expansion one finds 
\beq
T_\ell^{p^2} = - i 2 G_F V_{ud}^* F  \, m_\ell \, \bar{u}_L (p_\nu) \, v (p_\ell) ~ . 
\label{eq:Tp2} 
\eeq
$F$  can be identified  to lowest order with $F_\pi$ (and $F_K, F_\eta$).  
Setting $e=0$, to a given order ($p^{2n}$) in the purely "strong" chiral expansion, 
 the amplitude reads as above, with the replacement 
$F \to F_{\pi}^{(2n)}$, $F_\pi^{(2n)}$ being the pion decay constant to order $p^{2n}$.  
When considering the ratio of electron-to-muon decay rates  the pion decay constant drops
and one obtains  the well known expression:
\beq
R_{e/\mu}^{(0),(\pi)}  = \frac{m_e^2}{m_\mu^2}  \left(  \frac{m_\pi^2 - m_e^2}{m_\pi^2 - m_\mu^2}   \right)^2  ~.  
\label{eq:R0}
\eeq
Non-trivial corrections to Eq.~\ref{eq:R0} arise only when $e\neq0$, 
i.e. to order $e^2 p^{2n}$ in ChPT. 

Lorentz invariance implies that higher order contributions are proportional to the 
lowest order  amplitude,  and this allows one  to write to $O(e^2 p^4)$
\beq
\Gamma (\pi \to \ell \nu [\gamma]) = 
\Gamma^{(0)} (\pi \to \ell \nu) \times
\left[ 1 + 
2\, {\rm Re}  \left( r_\ell^{e^2p^2} + r_\ell^{e^2p^4} \right)  +
\delta_\ell^{e^2 p^2} +  \delta_\ell^{e^2 p^4} 
\right] ~ ,  
\eeq
where 
\beq
\Gamma^{(0)} (\pi \to \ell \nu) =  \frac{G_F^2 |V_{ud}|^2  F_\pi^2 }{4 \pi} \, 
m_\pi  \, m_\ell^2  \, \left(1 - \frac{m_\ell^2}{m_\pi^2} \right)^2 
\label{eq:gamma0}
\eeq
and 
\bea
r_\ell^{e^2 p^{2n}} &  = &  \frac{ T_\ell^{e^2 p^{2n}}}{T_\ell^{p^2}} 
\label{eq:rl} \\
\delta_\ell^{e^2 p^{2n}} &  = &  \frac{ \Gamma (\pi \to \ell \nu \gamma) \vert_{e^2 p^{2n}}
}{\Gamma^{(0)} (\pi \to \ell \nu) } 
\label{eq:deltae2p2n}
\eea
are respectively the corrections induced by virtual and real photon effects, whose sum is free
of infrared divergences.  Taking the ratio of electron and muon decay rates one obtains:
\bea
R_{e/\mu}^{(\pi)} &= & R_{e/\mu}^{(0),(\pi)}   
\, \Bigg[   1 + 
\Delta_{e^2 p^2}^{(\pi)} +   \Delta_{e^2 p^4}^{(\pi)}  +  ...  
\Bigg]   
\label{eq:Rmaster1}
\\  
\Delta_{e^2 p^{2n}}^{(\pi)} &=&   
 2\, {\rm Re}   \left( r_e^{e^2p^{2n}} - r_\mu^{e^2p^{2n}} \right) 
 + \left(\delta_e^{e^2p^{2n}} - \delta_\mu^{e^2p^{2n}} \right)  \
\label{eq:Rmaster2}
\eea
The  main feature emerging from  Eq.~\ref{eq:Rmaster2} is that only those diagrams 
that  depend in a non-trivial way on the lepton mass contribute to $R_{e/\mu}$. 
The diagrams leading to  $m_\ell$-independent  $r_{\ell}^{e^2p^{2n}}$ 
will drop when taking the difference of electron and muon amplitudes. 
This observation greatly reduces the number of diagrams to be calculated in the effective  theory. 
All the considerations presented in this section trivially extend to the case of leptonic decays 
of charged kaons ($K \to \ell \nu$).

\section{Electromagnetic corrections to (semi)-leptonic processes at low energy}
\label{sect:SM}

The appropriate theoretical framework for the analysis of electromagnetic 
effects in semileptonic kaon decays is a low-energy effective 
quantum field theory where the  asymptotic states consist of the 
pseudoscalar octet, the photon and the light leptons \cite{Knecht:1999ag}. 
The corresponding lowest-order effective Lagrangian is given by
\beqa \label{Leff}
\cL_{\rm eff} &=& \frac{F^2}{4} \; \langle u_\mu u^\mu + \chi_+\rangle +
e^2 F^4 Z \langle \Q_{\rm L}^{\rm em} \Q_{\rm R}^{\rm em}\rangle 
- \frac{1}{4} F_{\mu\nu} F^{\mu\nu}                \no \\
&& \mbox{} + \sum_\ell
[ \bar \ell (i \! \not\!\partial + e \! \not\!\!A - m_\ell)\ell +
\ol{\nu_{\ell \rm L}} \, i \! \not\!\partial \nu_{\ell \rm L}]  .
\eeqa
$F$ denotes the pion decay constant in the chiral limit and in the 
absence of electroweak interactions.
The low energy constant $Z \simeq 0.8$ can be determined by
mass splitting of  charged and neutral  pions.
The symbol $\langle \; \rangle$ denotes the trace in three-dimensional 
flavour space, and
\beq \label{umu}
u_\mu = i [u^\dg (\partial_\mu - i r_\mu)u - u
(\partial_\mu - i l_\mu)u^\dg] ~, 
\eeq
with the Goldstone modes collected in the field $u$:
\beq
u = \exp \left[ \frac{i \Phi}{\sqrt{2} F} \right]  \qquad \qquad \qquad 
\Phi =  \left[ \begin{array}{ccc}
\frac{\pi^0}{\sqrt{2}} + \frac{1}{\sqrt{6}} \eta_8   &     \pi^+  &  K^+  \\
\pi^-  &  - \frac{\pi^0}{\sqrt{2}} + \frac{1}{\sqrt{6} }\eta_8   &  K^0  \\
K^-    &   \bar{K}^0     &             - \frac{2}{\sqrt{6}} \eta_8   
\end{array} 
\right]~ .
\eeq
The photon field $A_\mu$ and the leptons $\ell,\nu_\ell$ ($\ell = e,\mu$)
are contained in (\ref{umu})
by adding appropriate terms to the usual external vector and axial-vector
sources $v_\mu$, $a_\mu$:
\beqa \label{sources}
l_\mu &=& v_\mu - a_\mu - e Q_{\rm L}^{\rm em} A_\mu + \sum_\ell
(\bar \ell \gamma_\mu \nu_{\ell \rm L} Q_{\rm L}^{\rm w} + \ol{\nu_{\ell 
\rm L}} 
\gamma_\mu \ell
Q_{\rm L}^{{\rm w}\dg})  , \no \\
r_\mu &=& v_\mu + a_\mu - e Q_{\rm R}^{\rm em} A_\mu  .
\eeqa
The $3 \times 3$ matrices $Q_{\rm L,R}^{\rm em}$, $Q_{\rm L}^{\rm w}$ are 
spurion 
fields.
At the end, one identifies 
$Q_{\rm L,R}^{\rm em}$ with the quark charge matrix
\beq \label{Qem}
Q^{\rm em} = \left[ \ba{ccc} 2/3 & 0 & 0 \\ 0 & -1/3 & 0 \\ 0 & 0 & -1/3 \ea
\right],
\eeq
whereas the weak spurion is taken at
\beq \label{Qw}
Q_{\rm L}^{\rm w} = - 2 \sqrt{2}\; G_{\rm F} \left[ \ba{ccc}
0 & V_{ud} & V_{us} \\ 0 & 0 & 0 \\ 0 & 0 & 0 \ea \right],
\eeq
where $G_{\rm F}$ is the Fermi coupling constant and $V_{ud}$, $V_{us}$ 
are Cabibbo-Kobayashi--Maskawa matrix elements.
For the construction of the effective Lagrangian it is also convenient
to define
\beq
\Q_{\rm L}^{\rm em,w} := u Q_{\rm L}^{\rm em,w} u^\dg  , \qquad
\Q_{\rm R}^{\rm em} := u^\dg Q_{\rm R}^{\rm em} u  .
\eeq
Explicit chiral symmetry breaking is included in
$\chi_+ = u^\dg \chi u^\dg + u \chi^\dg u$ where $\chi$ is 
proportional to the quark mass matrix: 
\beq \label{chi}
\chi = 2 B_0 \left[ \ba{ccc} m_u & 0 & 0 \\ 0 & m_d & 0 \\ 0 & 0 & m_s 
\ea
\right]~, 
\eeq
and the factor $B_0$ is related to the quark condensate in the chiral limit by 
$\langle 0 | \ol q q | 0 \rangle = -F^2 B_0$.

The local action at next-to-leading order involves the  
sum of three  terms, $\cL_{p^4} + \cL_{e^2 p^2}^{\rm str} + \cL_{e^2 p^2}^{\rm lept}$. 
The first one, $\cL_{p^4}$ includes the well-known Gasser-Leutwyler 
Lagrangian \cite{Gasser:1984gg}  in the presence of the generalized external sources 
introduced in (\ref{sources}), as well as a term from the Wess-Zumino-Witten functional 
that incorporates the effect of chiral anomalies~\cite{WZW}.
Here we quote only the  operators relevant to our analysis:
\beqa
\cL_{p^4} &\supset&
- iL_9 \  \langle f_+^{\mu\nu} u_\mu u_\nu\rangle + 
  \frac{L_{10}}{4} \,   \langle f_{+\mu\nu} f_+^{\mu\nu}
-  f_{-\mu\nu} f_-^{\mu\nu}\rangle \nn 
&& -\frac{iN_C}{48\pi^2} \varepsilon^{\mu\nu\alpha\beta} \langle \Sigma_\mu^L U^\dagger \partial_\nu r_\alpha U \ell_\beta 
- \Sigma_\mu^R U \partial_\nu \ell_\alpha U^\dagger r_\beta
 +\Sigma_\mu^L \ell_\nu \partial_\alpha \ell_\beta + \Sigma_\mu^L \partial_\nu \ell_\alpha \ell_\beta \rangle 
\label{L4}
\eeqa
with
\beqa
f_{\pm}^{\mu \nu} &=& u F_{\rm L}^{\mu \nu} u^{\dg} \pm 
                       u^{\dg} F_{\rm R}^{\mu \nu} u  , \nn
F_{\rm L}^{\mu \nu} &=& \partial^{\mu} l^{\nu} - \partial^{\nu} l^{\mu}
                  - i [l^\mu,l^\nu]  , \nn
F_{\rm R}^{\mu \nu} &=& \partial^{\mu} r^{\nu} - \partial^{\nu} r^{\mu}
                  - i [r^\mu,r^\nu]  , \nn 
U&=& u^2 , \nonumber \\
\Sigma_\mu^{L} &=& U^\dagger \partial_\mu U , \nn
\Sigma_\mu^R &=& U \partial_\mu U^\dagger .
\eeqa
The second term, $\cL_{e^2 p^2}^{\rm srt}$,  encodes the interaction of ultraviolet 
(UV) virtual photons with hadronic degrees of freedom 
\cite{urech,nr95,nr96}.  It contributes to the individual $P \to e  \nu$ and $P \to \mu \nu$, 
but leads to an $m_\ell$-independent $r_{\ell}^{e^2 p^2}$  so that it has no effect on $R_{e/\mu}$. 
The same argument applies to  $\cL_{e^2 p^2}^{\rm lept}$, which involves leptonic bilinears. 
Similarly, when inserted in one-loop purely mesonic graphs, these effective operators
contribute to $P \to e  \nu$ and $P \to \mu \nu$ to order $e^2 p^4$, but their contribution 
cancels in  $R_{e/\mu}$. 
Therefore, there is no need to report the full expression of  these effective lagrangians here.  

Finally, we shall see that a counterterm of $O(e^2 p^4)$ is needed in order to make 
$R_{e/\mu}$ finite to $O(e^2 p^4)$.  
While we have not constructed the most general ${\cal L}_{e^2 p^4}^{\rm lept}$, 
on the basis of power counting we  can conclude  that the same  combination 
of operators (and LECs)  contributes to both $R_{e/\mu}^{(K)}$  and $R_{e/\mu}^{(\pi)}$. 
This fact is also explicitly borne out in the matching calculation that we perform in 
Section~\ref{sect:matching}.

\begin{figure}[!t]
\centering
\begin{picture}(300,100)  
\put(120,25){\makebox(50,50){\epsfig{figure=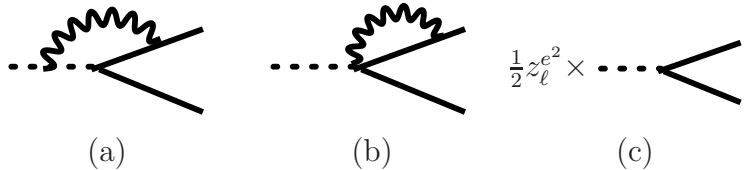,width=10cm}}}
\put(195,32){{
$\frac{1}{2}  z_{\ell}^{e^2} \times $}}
\put(40,0){(a)}
\put(140,0){(b)}
\put(240,0){(c)}
\end{picture}
\caption{ 
Diagrams contributing to $R_{e/\mu}$ to order $e^2 p^2$. 
Dashed lines indicate pseudoscalar mesons, solid lines leptons, and 
wavy lines photons.  
\label{fig:fig1}
}
\end{figure}

\section{Virtual-photon corrections: analysis}
\label{sect:analysis}

We work in Feynman  gauge, use dimensional regularization to deal with 
ultraviolet  (UV) divergences and an infinitesimal photon mass to deal with 
infrared (IR) divergences. 
We report the diagrams contributing to $R_{e/\mu}$ to $O(e^2 p^2)$
and $O(e^2 p^4)$ in Fig.~\ref{fig:fig1} and Figs.~\ref{fig:fig2}-\ref{fig:fig3}, respectively. 
At the order we work, we need the charged lepton  and pseudoscalar meson 
wavefunction renormalizations to one-loop accuracy.  We denote them by  
$Z_{\ell} = 1 + z_{\ell}^{e^2}$ (charged lepton) 
and  $Z_{\pi} = 1 + z_{\pi}^{p^2}  + z_{\pi}^{e^2}$  (pseudoscalar meson). 

To  order $e^2 p^2$  one has to consider only two 1PI diagrams and the 
effect of charged lepton wave-function renormalization (see Fig.~\ref{fig:fig1}). 
The resulting amplitude $T_\ell^{e^2 p^2}$~\cite{Knecht:1999ag} coincides with the point-like approximation of Ref.~\cite{Kinoshita:1959ha}.  Since this is  well known, we do not dwell 
further on it,  but we will simply report the result in the next section.  
The situation is more interesting to next-to-leading order. 
 
\subsection{Organizing the $O(e^2 p^4)$ diagrams}

\begin{figure}[!t]
\centering
\begin{picture}(300,300)  
\put(150,110){\makebox(50,50){\epsfig{figure=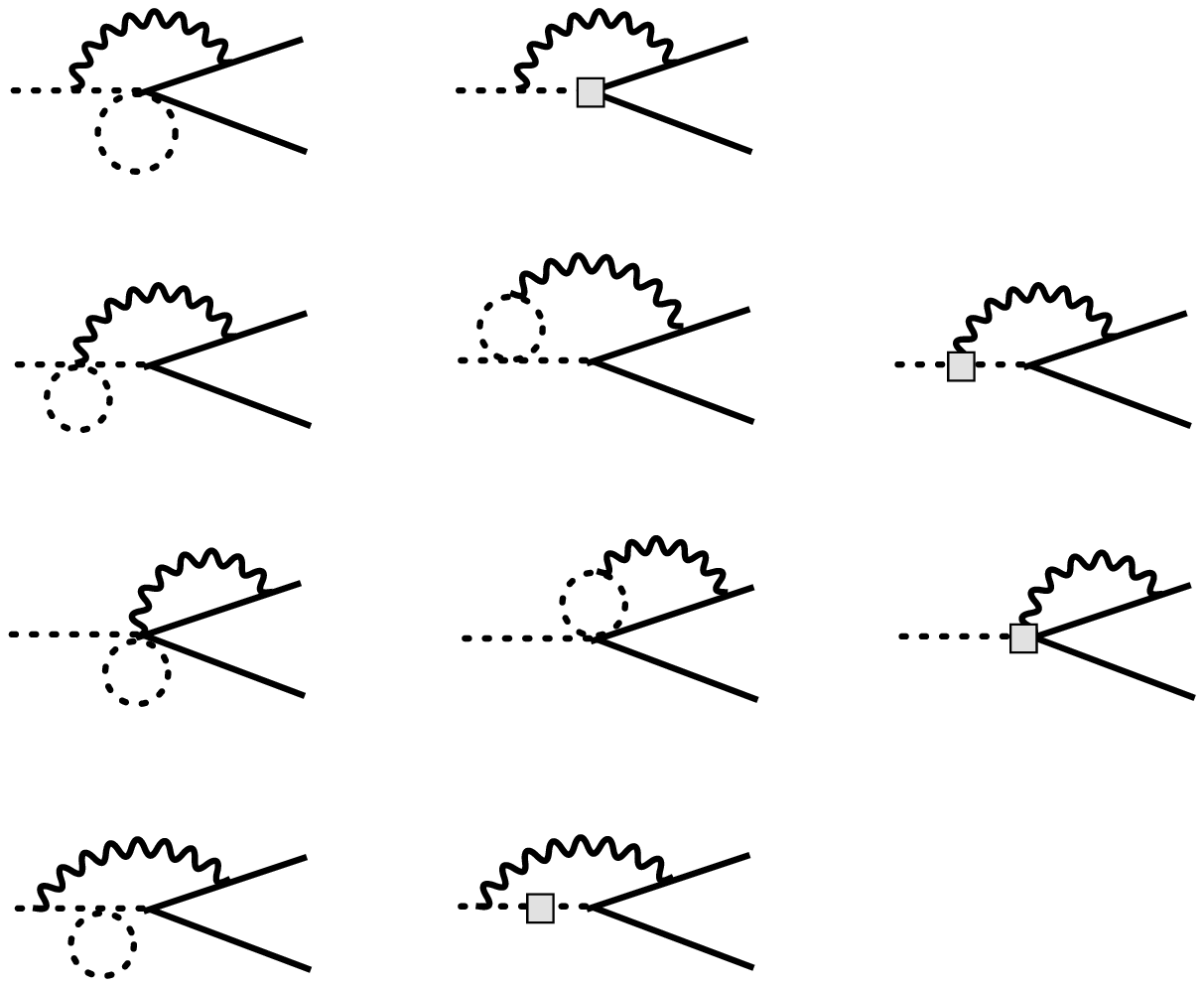,width=11cm}}}
\put(-50,23){{\small $(IV)$}}
\put(-50,93){{\small $(III)$}}
\put(-50,163){{\small $(II)$}}
\put(-50,233){{\small $(I)$}}
\end{picture}
\caption{ 
1PI diagrams contributing to $R_{e/\mu}$ to order $e^2 p^4$. 
Shaded squares indicate vertices from the $O(p^4)$ effective lagrangian. 
\label{fig:fig2}
}
\end{figure}

\begin{figure}[t]
\centering
\begin{picture}(300,175)  
\put(125,75){\makebox(50,50){\epsfig{figure=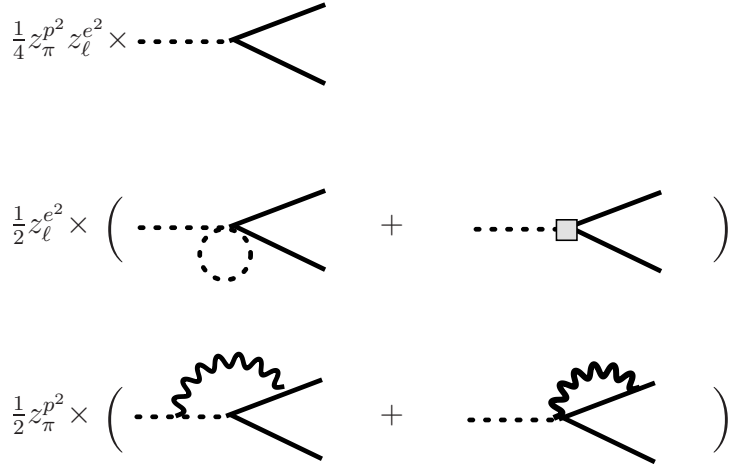,width=7cm}}}
\put(35,95){{  {\Huge  ( }  }}
\put(0,100){{ $\frac{1}{2}  z_{\ell}^{e^2} \times$  }}
\put(140,100){{ $+$ }}
\put(265,95){{  {\Huge  ) }  }}
\put(35,23){{  {\Huge  ( }  }}
\put(0,28){{ $\frac{1}{2}  z_{\pi}^{p^2} \times$  }}
\put(140,28){{ $+$ }}
\put(265,23){{  {\Huge  ) }  }}
\put(0,170){{ $\frac{1}{4}  z_{\pi}^{p^2}    z_{\ell}^{e^2}  \times$  }}
\end{picture}
\caption{ 
External leg corrections to  $R_{e/\mu}$ to order $e^2 p^4$. 
\label{fig:fig3}
}
\end{figure}

To $O(e^2 p^4)$  one has to consider (i)  two-loop graphs 
with vertices from the lowest order effective lagrangian and 
(ii) one-loop graphs with one insertion from the NLO lagrangian  ${\cal L}_{p^4}$ 
(we denote the latter vertices with shaded squares); 
(iii) a tree level diagram with insertion of a local operator of $O(e^2 p^4)$. 
In Fig.~\ref{fig:fig2}  we report all relevant 1PI  topologies: 
each $O(p^4)$ vertex receives contributions 
from several  $O(p^4)$ operators and all allowed mesons run in the internal loops.  
External leg corrections are depicted in Fig.~\ref{fig:fig3}.

\begin{figure}[t]
\centering
\begin{picture}(300,100)  
\put(125,25){\makebox(50,50){\epsfig{figure=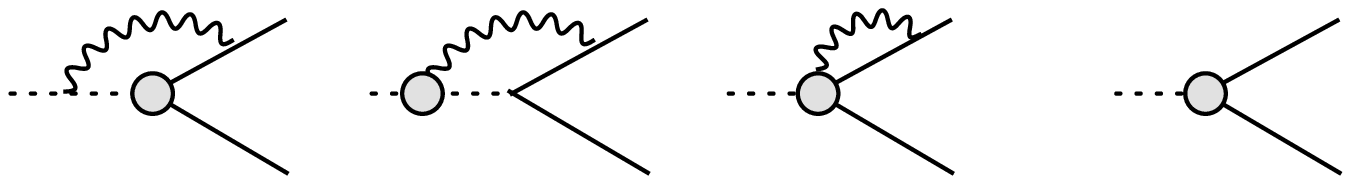,width=14cm}}}
\put(240,47){{ $\frac{1}{2}  z_{\ell}^{e^2}  \times$  }}
\put(300,5){(d) }
\put(190,5){(c)}
\put(90,5){(b)}
\put(-20,5){(a)}
\end{picture}
\caption{ 
Effective one-loop diagrams contributing to  $R_{e/\mu}$ to order $e^2 p^4$. 
The shaded circles represent the $O(p^4)$ contribution to d-dimensional  off-shell 
effective vertices. 
\label{fig:fig4}
}
\end{figure}

The self-energy insertion on the internal mesonic leg (class $(IV)$ in Fig.~\ref{fig:fig2}) is handled by observing that 
to $O(p^4)$  the self-energy reads  $\Sigma (p^2) =  A + B  p^2$ (with $A$ and $B$ momentum-independent)  and therefore 
\beq
\frac{i}{p^2 - m_0^2}   \left(- i \Sigma (p^2) \right)  \frac{i}{p^2 - m_0^2}  =  \left( Z - 1\right) \frac{i}{p^2 - m^2}  + 
(m^2 - m_0^2)  \frac{\partial}{\partial m_0^2}   \frac{i}{p^2 - m_0^2}  \ , 
\eeq
where $Z$ represents the on-shell wave-function renormalization,  $m_0$ is the $O(p^2)$ mass and $m$ is the physical 
$O(p^4)$ mass. 
With this result at hand, by re-grouping the diagrams of class $(IV)$ and external leg corrections 
with those of classes $(I)$, $(II)$, and $(III)$,  it is straightforward  to show that the inclusion of 
virtual corrections to  $O(e^2 p^4)$  amounts  to: 
\begin{itemize}
\item  using the physical $O(p^4)$ meson mass in the amplitude of $O(e^2 p^2)$; 
\item  calculating a set  of  "effective" one-loop diagrams with 
vertices given by appropriate off-shell form factors 
evaluated to $O(p^4)$ in d-dimensions.   These effective one-loop diagrams are shown in 
Fig.~\ref{fig:fig4}.   The shaded circles denote respectively: 
the d-dimensional $O(p^4)$  $\pi \ell \nu$  vertex 
(Fig.~\ref{fig:fig4}(a) and (d),  with off-shell pion and charged lepton in Fig.~\ref{fig:fig4}(a)); 
the d-dimensional $O(p^4)$  $\pi \pi \gamma$  vertex with the  photon and one pion off-shell
 (Fig.~\ref{fig:fig4}(b));
the d-dimensional $O(p^4)$  $ \pi  \ell \nu \gamma$  vertex with the photon and charged lepton off-shell (Fig.~\ref{fig:fig4} (c)). 
\end{itemize}
Within this approach one starts the calculation of genuine two-loop diagrams at a stage where 
the one-loop sub-divergences (generating non-local singularities) have already been subtracted. 
As we shall see,  another advantage  is that the non-local  O$(p^4)$ vertices  
admit a simple dispersive parameterization  that greatly simplifies the calculation. 

As seen from Fig.~\ref{fig:fig4},  the virtual photon contributions
can be divided into 1PI and external leg corrections. 
For the external leg corrections we find: 
\beq
T_\ell^{e^2 p^4} \Big|_{{\rm non}-1PI}  
= \frac{1}{2} z_{\ell}^{e^2} \  \left(\frac{F_\pi^{(4)}}{F} - 1 \right)   \times T_\ell^{p^2} \ , 
\label{eq:e2p4ext}
\eeq
where $F_\pi^{(4)}/F$ has to be evaluated in d-dimensions.    
The 1PI contribution can be written as the sum of the mass-renormalization in $T_{\ell}^{e^2 p^2}$ 
and a convolution:
\bea
T_\ell^{e^2 p^4} \Big|_{1PI} &=& 2 G_F V_{ud}^* e^2 F  \   \int  \frac{d^d q}{(2 \pi)^d} \,  
\frac{ \bar{u}_L (p_\nu) \gamma^\nu \left[  - (\slashed{p}_{\ell} - \slashed{q})  + m_{\ell}  \right] \gamma^\mu  v(p_\ell)}{
\left[q^2 - 2 q \cdot p_{\ell} + i \epsilon \right] \left[q^2 - m_\gamma^2  + i \epsilon \right ]} \ T^{V-A}_{\mu \nu} (p,q) 
\nonumber \\
&+& 
\left(  m_\pi^2 \big|_{p^4} - m_\pi^2 \big|_{p^2}  \right) \, 
 \frac{\partial}{\partial m_\pi^2}   \, T_{\ell}^{e^2 p^2} ~, 
\label{eq:convolution}
\eea
where
\beq
 T^{V-A}_{\mu \nu} =  \frac{1}{\sqrt{2} F} \, 
 \int d x \ e^{i q x + i W y} \   \langle 0  | T (J^{EM}_{\mu}  (x)  \, (V_\nu - A_\nu) (y)  | \pi^+ (p)  \rangle~, 
\label{eq:correlator1}
\eeq
with $V_\mu (A_\mu) = \bar{u} \gamma_\mu (\gamma_5) d$ and $W=p-q$.  
Lorentz invariance and Ward identities imply that 
$T^{V-A}_{\mu \nu} $ in turn can be decomposed as follows 
(see also~\cite{Bijnens:1992en}):~\footnote{In this work we use the 
convention $\epsilon_{0123}  = + 1$ for the Levi Civita symbol.}
\bea
\left( T^{V-A} \right)^{\mu \nu} (p,q) &=&  i V_1   \, \epsilon^{\mu \nu \alpha \beta} q_\alpha p_\beta  + 
\left[\frac{(2 p - q)^\mu (p - q)^\nu}{2 p \cdot q - q^2}  + g^{\mu \nu} \right]  \left( \frac{F_\pi^{(4)}}{F} - 1
\right) 
\nonumber \\
&-& A_1 \,  \left(  q \cdot p  g^{\mu \nu} - p^\mu q^\nu \right) 
- (A_2 - A_1)  \left( q^2 g^{\mu \nu}  - q^\mu q^\nu \right) 
\nonumber \\
&+& 
\left[
\frac{(2 p - q)^\mu (p - q)^\nu}{2 p \cdot q - q^2}   - 
\frac{q^\mu (p - q)^\nu}{q^2}  
\right]  \, \left( F_V^{\pi \pi} (q^2)  - 1\right)   
\nonumber \\
& -& A_3 \,   \left[
q \cdot p  \left( q^\mu p^\nu  - q^\mu q^\nu \right) + q^2 \left( 
p^\mu  q^\nu - p^\mu p^\nu 
\right)
\right]
\label{eq:correlator2}
\eea
The form factors $V_1, A_i$ depend in general on both $q^2$ and $W^2 = (p - q)^2$ 
and have to be evaluated to $O(p^4)$ in ChPT in d-dimensions~\footnote{To $O(p^4)$ the form factor $A_3$ vanishes.}.
The same applies to the pion form factor $F_V^{\pi \pi}(q^2) $ and decay constant $F_\pi$. 
The convolution integral generates a term proportional to $T_\ell^{e^2p^2} |_{1PI}$ as well as terms 
induced by $V_{1}$, $A_{1,2}$, and $F_V^{\pi \pi} - 1$.   
With obvious notation we can write
\bea
T_\ell^{e^2 p^4} \Big|_{1PI}  
&=& T_{V_1} + T_{A_1} +T_{A_2}  +T_{F_V}
\nonumber \\
& + &
\left(\frac{F_\pi^{(4)}}{F} - 1 \right)  \  T_\ell^{e^2 p^2} \Big|_{1PI}  
+
\left(  m_\pi^2 \big|_{p^4} - m_\pi^2 \big|_{p^2}  \right) \,  \frac{\partial}{\partial m_\pi^2}   \, T_{\ell}^{e^2 p^2}~. 
\label{eq:e2p41PI}
\eea
Combining Eqs.~\ref{eq:e2p41PI} and \ref{eq:e2p4ext} we then obtain:
\beq
T_\ell^{e^2 p^4} 
= T_{V_1} + T_{A_1} +T_{A_2} +T_{F_V} + 
\left(\frac{F_\pi^{(4)}}{F} - 1 \right)   \  T_\ell^{e^2 p^2}  
+ 
\left(  m_\pi^2 \big|_{p^4} - m_\pi^2 \big|_{p^2}  \right) \,  \frac{\partial}{\partial m_\pi^2}   \, T_{\ell}^{e^2 p^2} 
\label{eq:e2p4full}
\eeq
The  effect of the the last two terms in Eq.~\ref{eq:e2p4full} is taken into account by 
simply using  the physical pion mass  and decay constant to $O(p^4)$ in $T_{\ell}^{e^2 p^2}$.
The remaining terms provide a genuine shift to the invariant amplitude. 
In order to calculate such a shift,  we need to: 
\begin{itemize}
\item[(i)] Work out 
the relevant form  factors $V_1, A_{1,2}, F_V^{\pi \pi}$  to $O(p^4)$ (one-loop) in d-dimensions. 
\item[(ii)]  Insert them in the convolution representation of Eq.~\ref{eq:convolution}  
and calculate the resulting integrals. 
\end{itemize}
In the following subsections we report the results of these steps.

\subsection{Form factors in d-dimensions} 
\label{sect:ff}
We work in $d$ dimension with $d = 4 + 2 w$~\cite{Bijnens:1997vq,Gasser:1998qt}. 
The relevant form factors to $O(p^4)$ read:
\bea
V_1 &=& - \frac{N_C}{24 \pi^2 F^2} 
\label{eq:V1} \\
A_1 &=& - \frac{4 \left( L_9 + L_{10}\right)}{F^2}
\label{eq:A1} \\
A_2 &=& - 2 \frac{\left( F_V^{\pi \pi} (q^2) - 1 \right) }{q^2}
\label{eq:A2} \\
F_V^{\pi \pi} (q^2)  &=& 1 + 2 \, H_{\pi \pi} (q^2)  + H_{KK} (q^2) 
\label{eq:FV}  
\eea
The loop function $H_{aa} (q^2)$~\cite{Gasser:1984gg} reads
\beq
F^2  H_{aa} (q^2) = q^2 \left[ 
\frac{A(m_a^2)}{m_a^2} \frac{d-2}{8(d-1)}  + \frac{2}{3} L_9  \right]
+ \frac{q^2 - 4 m_a^2}{4 (d-1)}  \, \bar{J}^{aa}(q^2) ~,  
\eeq
with
\bea
A(m^2)& \equiv&  - i \int \frac{d^d k}{(2 \pi)^d} \,  \frac{1}{k^2 - m^2 + i \epsilon}  = 
- \frac{m^{2 + 2 w}}{(4 \pi)^{2 +w}} \, \Gamma(- 1 - w) \\
\bar{J}^{aa}(q^2)  & = & J^{aa} (q^2) - J^{aa}(0) \\
J^{aa} (q^2) &\equiv  & 
- i \int \frac{d^d k}{(2 \pi)^d} \,  \frac{1}{\left[k^2 - m_a^2 + i \epsilon \right] \ \left[(k - q)^2- m_a^2 + i \epsilon
\right]} \nonumber \\
&=& \frac{1}{(4 \pi)^{2 + w}} \Gamma(-w) \, \int_0^1 \, dx   \left[
m_a^2 - q^2 \,  x (1 -x) \right]^w 
\eea
The function $J^{aa}(q^2)$ admits a dispersive 
representation~\cite{Bijnens:1997vq,Gasser:1998qt} in d-dimensions, which 
proves very useful in the evaluation of genuine two-loop contributions:
\bea
J^{aa}(q^2) &=&   m_a^{2w}  \int_{4 m_a^2}^{\infty} \, \left[ d \sigma \right] \frac{1}{\sigma - q^2} \\
\left[ d \sigma  \right] & = &  \frac{d \sigma}{(4 \pi)^{2+w}} \, \frac{\Gamma \left(\frac{3}{2} \right)}{\Gamma 
\left(\frac{3}{2} + w\right)} \, \left(\frac{\sigma}{4 m_a^2} - 1 \right)^w \left( 1- \frac{4 m_a^2}{\sigma} 
\right)^{1/2} 
\label{eq:Jdisp1}
\eea

\subsection{"Effective" one-loop diagrams}
\label{sect:basic}

The 1PI contributions  $T_{V_1} , T_{A_1} , T_{A_2} , T_{F_V} $ can be written as the  
convolution of a known kernel with the d-dimensional form factors 
$V_{1}, A_{1,2}, F_V^{\pi\pi}$.  
It is simple to check that  $T_{V_1} , T_{A_1} , T_{A_2} , T_{F_V} $ are IR finite, 
so we set $m_\gamma=0$. 
Upon inserting the $O(p^4)$  ChPT form factors into
Eqs.~\ref{eq:correlator2} and \ref{eq:convolution},  we obtain:
\bea 
T_{V_1}& =&  T_\ell^{p^2} \, e^2  V_1  \,  \frac{A(m_\ell^2)}{2 (d-1) m_\ell^2} 
\, \left[ (4 + d) m_\ell^2 - (d -2)  m_\pi^2 \right] 
\label{eq:TV1}
\\
T_{A_1}& =& - T_\ell^{p^2} \, e^2  A_1  \,  \frac{A(m_\ell^2)}{4 (d-1) m_\ell^2} 
\, \left[ (3 d^2 - 6 d + 4) m_\ell^2 - (d -2)^2  m_\pi^2 \right] 
\label{eq:TA1}
\\
T_{A_2}& =& - 2 \, T_\ell^{p^2} \, e^2     \Bigg\{ 
d \, \left( 2  a_{\pi \pi} + a_{KK} \right)  A (m_\ell^2)  
\nonumber \\ 
&+&   \frac{2 b_{\pi \pi}}{i}  \left[
2 I_{2}^{(\ell) \pi \pi} + I_{2}^{(\ell) KK} -   8 m_\pi^2 I_{1}^{(\ell) \pi \pi} 
-   4 m_K^2 I_{1}^{(\ell) KK} 
\right]
\nonumber \\
&+&    \frac{b_{\pi \pi}}{i m_\ell^2}  \left(1 - {d \over 2} \right)  \,  \Big[
2 \left( I_{4}^{\pi \pi}  - I_5^{\pi \pi}  \right) + \left( I_{4}^{KK}  - I_5^{KK}  \right)
\nonumber \\
&-&   
8 m_\pi^2 \left( I_{3}^{\pi \pi}    - I_{2}^{(\ell) \pi \pi}   \right)
-   4 m_K^2 \left( I_{3}^{KK}    - I_{2}^{(\ell) KK}   \right)
\Big]  \Bigg\}
\label{eq:TA2}
\\
T_{F_V}& =&  2 \, T_\ell^{p^2} \, e^2     \Bigg\{ 
\left( 2  a_{\pi \pi} + a_{KK} \right)  \, 
\frac{m_\pi^2  A(m_\pi^2)  - m_\ell^2 A(m_\ell^2) }{m_\pi^2 - m_\ell^2}   
\nonumber \\
&+ & 
\frac{m_\pi^2}{m_\pi^2 - m_\ell^2} \ \frac{b_{\pi \pi}}{i} \left(
2 I_{2}^{(\ell) \pi \pi} + I_{2}^{(\ell) KK} -   8 m_\pi^2 I_{1}^{(\ell) \pi \pi} 
-   4 m_K^2 I_{1}^{(\ell) KK} \right) 
\nonumber \\
& -& 
\frac{m_\ell^2}{m_\pi^2 - m_\ell^2}   
 \frac{b_{\pi \pi}}{i} \left(
2 I_{2}^{(\pi) \pi \pi} + I_{2}^{(\pi) KK} -   8 m_\pi^2 I_{1}^{(\pi) \pi \pi} 
-   4 m_K^2 I_{1}^{(\pi) KK} \right) 
\nonumber \\
& +& 
(m_\pi^2 + m_\ell^2)  
\frac{b_{\pi \pi}}{i} \left(
2 T_{2}^{\pi \pi} + T_{2}^{KK} -   8 m_\pi^2 T_{1}^{\pi \pi} 
-   4 m_K^2 T_{1}^{KK} \right) 
\Bigg\}
\label{eq:TFV}
\eea
In the above expressions we have used the definitions:
\bea
a_{\pi \pi}& =& \frac{1}{F^2} \left( 
\frac{A(m_\pi^2)}{m_\pi^2} \frac{d-2}{8(d-1)}  + \frac{2}{3} L_9 
\right) 
\\
a_{KK} &=& \frac{1}{F^2} \left( 
\frac{A(m_K^2)}{m_K^2} \frac{d-2}{8(d-1)}  + \frac{2}{3} L_9 
\right) 
\\
b_{\pi \pi} &=& \frac{1}{4 (d-1) \, F^2} 
\eea
which come from  the decomposition 
$H_{mm}(q^2) = a_{mm} q^2 + b_{mm} (q^2  - 4 m_m^2) \bar{J}^{mm}(q^2)$.
Moreover, the building-block  two loop integrals  are defined as follows: 
\bea
I_1^{(\ell)aa} &=&    \int \frac{d^d q}{(2 \pi)^d} \ 
\frac{ \bar{J}^{aa}(q^2) }{q^2 \left(q^2 - 2 q \cdot  p_\ell \right)}  
\\
I_1^{(\pi)aa} &=&    \int \frac{d^d q}{(2 \pi)^d} \ 
\frac{ \bar{J}^{aa}(q^2) }{q^2 \left(q^2 - 2 q \cdot  p \right)}  
\\
I_2^{(\ell) aa} &=&    \int \frac{d^d q}{(2 \pi)^d} \ 
\frac{ \bar{J}^{aa}(q^2) }{q^2 - 2 q \cdot  p_\ell }  
\\
I_2^{(\pi) aa} &=&    \int \frac{d^d q}{(2 \pi)^d} \ 
\frac{ \bar{J}^{aa}(q^2) }{q^2 - 2 q \cdot  p }  
\\
I_3^{aa} &=&    \int \frac{d^d q}{(2 \pi)^d} \ 
\frac{ \bar{J}^{aa}(q^2) }{q^2}  
\\
I_4^{aa} &=&    \int \frac{d^d q}{(2 \pi)^d} \  \bar{J}^{aa}(q^2) 
\\
I_5^{aa} &=&    \int \frac{d^d q}{(2 \pi)^d} \ 
\frac{ q^2  \, \bar{J}^{aa}(q^2) }{q^2 - 2 q \cdot  p_\ell }  
\\
T_1^{aa} &=&    \int \frac{d^d q}{(2 \pi)^d} \ 
\frac{ \bar{J}^{aa}(q^2) }{q^2 \left(q^2 - 2 q \cdot  p_\ell \right) \left(q^2 - 2 q \cdot  p \right)}  
\\
T_2^{aa} &=&    \int \frac{d^d q}{(2 \pi)^d} \ 
\frac{ \bar{J}^{aa}(q^2) }{\left(q^2 - 2 q \cdot  p_\ell \right) \left(q^2 - 2 q \cdot  p \right)}  ~ .
\eea
The evaluation of these integrals can be done analytically and is reported in  
Appendix~\ref{sect:technicalities}. 

\subsection{$K$ decays}
\label{sect:K}
The procedure outlined above remains true for the analysis of $K \to \ell \nu$.  
In the convolution kernel 
one has to simply replace $p \to p_K$ ($p^2=m_\pi^2 \to p_K^2=m_K^2$). 
The form factors $V_1$ and $A_1$ remain unchanged, while in $A_2$ one has to 
replace $F_V^{\pi \pi}(q^2) \to F_V^{KK}(q^2) =   1 + 2 H_{KK} (q^2) + H_{\pi \pi}(q^2)$, 
which again amounts to the interchange $m_\pi \leftrightarrow m_K$. 
As a consequence, the full result for $T_{\ell}^{e^2 p^4}  (K \to \ell \nu)$ can be obtained from 
the pion case by interchanging everywhere  $m_\pi$ with $m_K$.


\section{Virtual-photon corrections: results}
\label{sect:results}

We collect here the results for $r_{\ell}^{e^2 p^{2n}} =   T_{\ell}^{e^2 p^{2n}}/T_\ell^{p^2}$. 
Since $R_{e/\mu} \propto  r_{e}^{e^2 p^{2n}} - r_{\mu}^{e^2 p^{2n}}$, we 
systematically neglect $m_\ell$-independent contributions to $r_\ell^{e^2 p^{2n}}$  
that would drop in the difference. 
We also introduce the notation:
\beq
z_\ell \equiv \left( \frac{m_\ell}{m_\pi}\right)^2  \qquad  
z_\gamma \equiv \left( \frac{m_\gamma}{m_\pi}\right)^2  \qquad 
\tilde{z}_\ell  \equiv \left( \frac{m_\ell}{m_K}\right)^2 \qquad 
\tilde{z}_\pi  \equiv \left( \frac{m_\pi}{m_K}\right)^2 
~ .
\eeq

\subsection{Leading order: $r_{\ell}^{e^2 p^2}$} 
The one loop virtual photon contributions read~\cite{Knecht:1999ag}:
\bea
r_{\ell}^{e^2 p^2} &= & - \frac{\alpha}{2 \pi}   \, \log  \sqrt{z_\gamma}  
\  \left[ \frac{1 + z_\ell}{1 - z_\ell}  \,  \log z_\ell \right] 
\nonumber \\
&+&  \frac{\alpha}{4 \pi}   
\left[ 
\frac{7}{2} \log \frac{m_\ell^2}{\mu^2}   + \log z_\ell - \frac{2}{1 - z_\ell}  \log z_\ell  + 
\frac{1}{2} \frac{1 + z_\ell}{1 - z_\ell} \left( \log z_\ell \right)^2
\right]~.
\label{eq:virtuale2p2}
\eea
Note that the dependence on the renormalization scale $\mu$ drops in $R_{e/\mu}$. 
Moreover, the dependence on the IR regulator $m_\gamma$ disappears once the 
effect of real photon emission is included. 

\subsection{Next to leading order: $r_{\ell}^{e^2 p^4}$} 

Using the notation $\bar{L}_{9,10}  \equiv   (4 \pi)^2  \ L_{9,10}^r (\mu) $  and  
$\ell_{\alpha} = \log \frac{m_{\alpha}^2}{\mu^2}$ with $ \alpha= \pi,K,\ell$,   
we find the following expressions for the divergent and finite parts of the $O(e^2 p^4)$ amplitudes:
%
\bea
r_\ell^{e^2 p^4} \Big|_{V_1} &=& 
e^2   \frac{(\mu c)^{4w}}{(4 \pi)^4} \,     \frac{1}{w} \  \frac{8  N_C }{9} \frac{m_\ell^2}{F^2} \  + \ 
\frac{\alpha}{4 \pi} \, V_1 \, 
\left[
\frac{m_\pi^2}{3} \ell_\ell
 + m_\ell^2 
\left(\frac{5}{9} - \frac{4}{3}   \ell_\ell  
\right)
\right] 
\\
r_\ell^{e^2 p^4} \Big|_{A_1} &=& 
- e^2 \frac{(\mu c)^{4w}}{(4 \pi)^4} \,  \frac{1}{w} \, \frac{28}{3} \left( 
\bar{L}_{9} + \bar{L}_{10} 
\right) 
\frac{m_\ell^2}{F^2}   \ + \
\frac{\alpha}{4 \pi} \, A_1  
\left[
- \frac{m_\pi^2}{3} \ell_\ell
+ m_\ell^2  \left(\frac{13}{9} + \frac{7}{3}  \ell_\ell
\right)
\right] 
\\
r_\ell^{e^2 p^4} \Big|_{A_2} &=& 
e^2  \frac{(\mu c)^{4w}}{(4 \pi)^4} \, 
\left[ 
 \frac{1}{w^2}   + \frac{1}{w} \ 
\left( \frac{3}{2} + 16 \bar{L}_9   \right)  \right] \,    \frac{m_\ell^2}{F^2} 
 \nonumber \\
&+&
\frac{\alpha}{4 \pi} \frac{m_\ell^2}{(4 \pi F)^2} \, 
\Bigg\{  \left[  
\frac{13}{9}  + 8 \, \bar{L}_9  + \frac{16}{9} \left( \ell_\pi + \frac{1}{2} \ell_K \right)
+ \frac{2}{3} \left( \ell_\pi^2 + \frac{1}{2} \ell_K^2  \right) \right] \nonumber \\
&+&  4 \left(4 \bar{L}_9 - \frac{1}{6}  - \frac{1}{3} \ell_\pi - \frac{1}{6} \ell_K \right)  \, \ell_\ell  
- \frac{8}{9} \log ( z_\ell \sqrt{\tilde{z}_\ell}) + f_1 (z_\ell)  + \frac{1}{2} \, f_1 (\tilde{z}_\ell) 
\Bigg\}
\eea
\bea
r_\ell^{e^2 p^4} \Big|_{F_V} &=&
- e^2  \frac{(\mu c)^{4w}}{(4 \pi)^4} \, 
\left[
\frac{1}{4\, w^2} 
+  \frac{1}{w}  
\left(  \frac{5}{12}  +  4 \bar{L}_9 \right)
\right] \,  \frac{m_\ell^2}{F^2} 
\nonumber \\
&+&  \frac{\alpha}{4 \pi} \frac{m_\ell^2}{(4 \pi F)^2} \, 
\Bigg\{ \left[ - \frac{19}{36} -  \left( \frac{1}{2} + 4 \bar{L}_9 \right)  \, \ell_\pi + \frac{1}{6} 
\ell_\pi^2 + \frac{1}{6}  \ell_\pi \ell_K - \frac{1}{3} \ell_K - \frac{1}{12} \ell_K^2 \right]
\nonumber \\
&+& \frac{z_\ell}{1 - z_\ell} \, \log z_\ell    
 \left(4 \bar{L}_9 - \frac{1}{6}  - \frac{1}{3} \ell_\pi - \frac{1}{6} \ell_K \right)
 +  f_2 (z_\ell) +  f_3 (\tilde{z}_\ell, \tilde{z}_\pi) 
\Bigg\}~.
\eea
In terms of the building block functions $\tilde{E}_n (x)$, $\tilde{R}_n(x)$, 
$T^{\pi \pi} (x)$, and $T^{KK}(x,y)$    
defined in Appendix~\ref{sect:technicalities}  (Eqs.~\ref{eq:Etilde} and \ref{eq:useful1}-\ref{eq:useful}),  
the finite  functions  $f_{1,2,3}$  read:
\bea
f_1 (x)   &=&  
\frac{97}{54}   + \frac{4}{3} \,  
\left( 4 
\left( \tilde{R}_0  (x) + \frac{1}{6} \log x \right)
-  \tilde{R}_1 (x)
- 4 \tilde{R}_2 (x)
 + \tilde{R}_3 (x) 
  \right)     
 \nonumber \\
&+&  \frac{1}{3}  
\left( - 9 \tilde{E}_0 (x)
+ 6  \tilde{E}_1 (x)
+  8  \tilde{E}_2 (x)
- 6  \tilde{E}_3 (x)
+    \tilde{E}_4 (x)
 \right) 
\label{eq:x2barapp}
\\
f_2 (x)  &=&  
\frac{490 + 3 (147 - 32 \pi^2)}{108}   + \frac{1}{3} T^{\pi \pi}(x) 
 + \frac{1}{3} \frac{T^{\pi \pi}(x) - T^{\pi\pi}(0)}{x}  
 + \tilde{R}_2(x) -  \tilde{R}_0 (x) 
 \nonumber \\
&+&  \frac{1}{1 - x} \Bigg[ x \, 
\left(  \tilde{R}_2(x) -  \tilde{R}_0 (x) \right)   + \frac{1}{3} x\, 
\left( \tilde{E}_2(x) -  \tilde{E}_0 (x) \right)   
\nonumber \\
&+& \!\!\!\!  x\,   \left( \frac{540 + 3 (147 - 32 \pi^2)}{108}  \right) 
+ \frac{1}{3} \left( 2 (\tilde{E}_0 (x) - \tilde{E}_2 (x) )
+ \tilde{E}_3 (x) -  \tilde{E}_1 (x)  \right) \Bigg]
\\
f_3 (x,y)   &=&  - \frac{25}{108} 
+ \frac{1}{6} T^{KK}(x,y) 
+ \frac{1}{6} \frac{T^{KK}(x,y) - T^{KK}(0,y)}{x/y}  
\nonumber \\
&+& \frac{1}{6} \left(4 - y \right) 
\left(\tilde{R}_2 (x) - \tilde{R}_0 (x) \right)  
+ \frac{1}{3} \frac{y - 2}{y - x}     
\left(\tilde{E}_2 (y) - \tilde{E}_0 (y) \right)  
\nonumber \\
&+& \frac{1}{6} \frac{1}{1 - x/y}  
\Bigg[
\tilde{E}_3 (x) -\tilde{E}_1 (x)
- \tilde{E}_3 (y) +\tilde{E}_1 (y)
\nonumber \\
&+& x/y  \left( 4 - y \right) 
\left(\tilde{R}_2 (x) - \tilde{R}_0 (x) \right)  + 
(x/y - 2) \left(\tilde{E}_2 (x) - \tilde{E}_0 (x) \right)  
\Bigg]
\eea
Note that the functions $f_{1,2} (x)$ and $f_{3}(x,y)$ 
are non-singular for $x \to 0$ (corresponding to $m_\ell \to 0$). 

In order to make the $O(e^2 p^4)$ amplitude UV finite we introduce in the EFT a local counterterm.
By power counting such a term cannot distinguish $K$ and $\pi$ decays. 
Its contribution to the amplitude is:
\beq
r_\ell^{e^2 p^4} \Big|_{CT}  =  e^2 \,  \frac{m_\ell^2}{F^2} \,  \frac{(\mu c)^{4w}}{(4 \pi)^4} \, 
\left[\frac{d_2}{w^2}  + \frac{d_1^{(0)} + d_1^{(L)} (\mu) }{w}   + r_{CT} (\mu)  \right]
\label{eq:rCT}
\eeq
with 
\bea
d_2 &=&  - \frac{3}{4} \\
d_1^{(0)} &=&  - \frac{15}{4} \\
d_1^{(L)} (\mu)  &=&  -\frac{8}{3}  \bar{L}_9 (\mu) 
 +\frac{28}{3}  \bar{L}_{10} (\mu) ~.
\eea
The finite coupling $r_{CT} (\mu)$ satisfies the following renormalization group equation:
\beq
\mu \, \frac{d}{d \mu}  r_{CT} (\mu) =   - \left(  4 \, d_1^{(0)}  + 2 \, d_1^{(L)} (\mu)  \right)~ . 
\eeq
%


\section{Matching} 
\label{sect:matching} 

\subsection{Strategy}

Within  ChPT, the loop  calculation of $T_{\ell}^{e^2 p^4}$  produces  an ultraviolet divergence 
proportional to $(\alpha/\pi) m_\ell^2/(4 \pi F)^2$,  indicating the need to introduce a 
local operator of  $O(e^2 p^4)$, with an associated low-energy coupling.  
While the divergent part of the effective coupling is fully determined  by 
our loop calculation,  in order to estimate its finite part one needs to go beyond the 
low-energy effective theory and use information on the underlying QCD dynamics.   

In full generality, the $O(\alpha)$  virtual-photon correction to the  $\pi \to \ell \nu$  amplitude is given by a sum of contributions   that share the following convolution structure:
\beq
T_\ell^{e^2p^4} \Big\vert_{QCD}  = 
\int \frac{d^d q}{(2 \pi)^d}  \,  K  (q, p, p_e) \,  \Pi_{QCD} (q^2, W^2)   \  \Bigg\vert_{e^2 p^4}  ~ , 
\label{eq:match1}
\eeq
where $K $ is a known kernel,  $\Pi_{QCD}$  stands for one of the invariant form 
factors appearing in Eq.~\ref{eq:correlator2}, and  one has to   
expand the r.h.s.  up to  $O(e^2 p^4)$ in the chiral power counting. 
In the framework of the low-energy effective theory, when  calculating  $T_\ell^{e^2 p^4} $
 we use the $O(p^4)$  ChPT representation for the form factors, $\Pi_{QCD} \to \Pi_{ChPT}^{p^4}$ in Eq.~\ref{eq:match1}.  
While this representation is valid at scales below $m_\rho$
(and generates the correct single- and double-logs upon integration)  
it leads to the incorrect UV behavior of the integrand in \ref{eq:match1}, 
which is dictated by the Operator Product Expansion (OPE) for the 
$\langle V V P \rangle$ and  $\langle V A P \rangle$ correlators.  
As anticipated, this forces the introduction of a  local operator of $O(e^2 p^4)$ 
whose finite coupling is a priori unknown, so that: 
\beq
T_\ell^{e^2 p^4} \Big\vert_{ChPT}   = 
\int \frac{d^d q}{(2 \pi)^d}  \,  K  (q, p, p_e) \,  \Pi_{ChPT}^{p^4}  (q^2, W^2)  
+  T_{\ell}^{e^2 p^4, CT} ~ .  
\label{eq:match2}
\eeq
The physical matching condition $T_\ell^{e^2 p^4}\vert_{ChPT} = T_\ell^{e^2 p^4}\vert_{QCD}$
in principle allows one to determine the finite part of the counterterm.  From   
the above discussion it is evident that 
the counterterm arises from the UV region in the convolution 
of Eq.~\ref{eq:match1}, so  in order to estimate it 
we need a suitable representation of the correlators  which is valid for momenta 
beyond the chiral regime.   
This poses a complex non-perturbative problem that we are not able to solve within full QCD.  

The problem becomes tractable if we  work within the context of  a truncated 
version of large-$N_C$  QCD,  in which we replace  
$ \Pi_{QCD} \to \Pi_{QCD_\infty}$  and $\Pi_{ChPT} \to \Pi_{ChPT_\infty}$. 
In this framework we approximate the  
full QCD correlators by meromorphic functions, i.e. 
we assume that the correlators are saturated by the exchange of a {\it finite}  number 
of narrow resonances (at large $N_C$ one would have an infinite number of resonances).    
The relevant resonance couplings are fixed by requiring that the correlators obey 
suitable sets of QCD short-distance constraints~\cite{derafael}  
(see discussion in the next section for an assessment of the model-dependence). 
Correspondingly, in the chiral effective theory the correlators are obtained by 
considering  only tree-level diagrams involving 
Goldstone modes,  with the couplings of higher order operators  
 (in our case $L_9$ and $L_{10}$)  consistently determined by integrating out the resonance fields.    
In this framework we are able to perform all integrations analytically and 
we determine the local coupling by the matching condition:
\bea
 T_{\ell}^{e^2 p^4, CT}  &=& 
 \int \frac{d^d q}{(2 \pi)^d}  \,  K  (q, p, p_e) \,  \Pi_{QCD_{\infty}} (q^2, W^2)   \  \Bigg\vert_{e^2 p^4}  
\nonumber \\
  &-& \int \frac{d^d q}{(2 \pi)^d}  \,  K  (q, p, p_e) \,  \Pi_{ChPT_{\infty}}^{p^4}  (q^2, W^2)\,. 
\label{eq:match3} 
\eea
Note that since we are using the large-$N_C$ representations for the QCD and ChPT form factors ($ \Pi_{QCD_{\infty}}$ and $\Pi_{ChPT_{\infty}}^{p^4}$), our matching procedure is going to 
miss  corrections to  $c_3^{CT}(\mu )$  sub-leading   
in the $1/N_C$ expansion, which are responsible for the "double-log" scale 
dependence of the counterterm. 

\subsection{Meromorphic approximation for  the form factors} 
 
In  order to  implement the program described above, we need a suitable 
representation of the hadronic correlator of Eq.~\ref{eq:correlator2}, to be used in the 
convolution integral of Eq.~\ref{eq:convolution}.        
In Refs.~\cite{VVPrefs,mouss97,kn01,vcetal04,jorge} one can find analysis of
the $\langle VVP \rangle$ and   $\langle VAP \rangle$ Green Functions,
describing them in terms of simple meromorphic functions that respect the 
constraints imposed at low-momentum transfer by chiral symmetry  
and at high momentum-transfer by the OPE. 
For correlators that are order parameters of spontaneous 
chiral symmetry breaking, such as $\langle VVP \rangle$ and   $\langle VAP \rangle$,  
this is a sensible approximation  well supported by a number of studies~\cite{derafael}. 

Using the LSZ reduction formula, it is simple to extract  the form factors from the correlators 
of  Refs.~\cite{kn01,vcetal04}.
Denoting $M_V$ and $M_A$ the masses of vector and axial-vector meson resonances,  
and using $W=p-q$, we find:
\bea
 V_1 (q^2, W^2)  &=& \frac{1}{6}     \frac{2 (q^2 - q \cdot p)  - \frac{N_C  \, M_V^4}{4  \pi^2 F^2}}{
 (q^2 - M_V^2)  (W^2 - M_V^2)} 
 \\
 A_1 (q^2, W^2)  &=&  \frac{M_V^2 - M_A^2 - b_2 q^2  - b_3 W^2}{(q^2 - M_V^2) (W^2 - M_A^2)}  
 \\
 A_2 (q^2, W^2)  &=&  \frac{ -2 \,  M_A^2 - d_2  W^2}{(q^2 - M_V^2) (W^2 - M_A^2)}  
\\
A_3 (q^2, W^2)  &=& 
 - \frac{2 + d_2}{(q^2 - M_V^2) (W^2 - M_A^2)}  
\\
F_V( q^2) &=&  \frac{M_V^2}{M_V^2 - q^2} ~ . 
\eea
To leading order in powers of  $q^2$ and $W^2$,  the above results reproduce the 
ChPT results to $O(p^4)$,  Eqs.~\ref{eq:V1}, \ref{eq:A1}, \ref{eq:A2}, \ref{eq:FV},  
provided one identifies the low-energy constants with  their 
resonance-saturated values  $L_9 \to  F^2/(2 M_V^2)$ and 
$L_{10} \to -F^2/4  (1/M_V^2 + 1/M_A^2)$, and provided one neglects the chiral loops. Note that $A_3(q^2,W^2)$ is not relevant for our matching procedure, since it starts to contribute to our amplitude to $O(e^2p^6)$.
The dimensionless constants $b_{2,3}$ and $d_2$~\cite{vcetal04} 
are {\it a priori} unknnown and can be fixed by imposing constraints on the  asymptotic behavior 
of the Green Functions.     Different results exist in the literature, corresponding   
to different choices of the resonance content of the meromorphic ansatz and consequently 
different  sets of QCD short-distance constraints.   
These different choices will allow us to quantify at least in part 
the model-dependence  of the final answer.  
Let us briefly discuss the two choices:
\begin{itemize}
\item[1.] The authors of Refs.~\cite{mouss97,kn01} include in their hadronic ansatz for the 
$\langle V A P \rangle$ correlator  
only the lowest lying  V and A resonances and after imposing short-distance constraints they 
find 
\beq
b_2 = \frac{1}{2}~, \quad  b_3= -\frac{1}{2}~, \quad  d_2 = -1~. 
\eeq

\item[2.] On the other hand, the authors of Ref.~\cite{vcetal04} include also one multiplet of pseudoscalar 
(P) resonances in the truncated spectrum.  After imposing a larger set of  
short distance constraints they find $b_2 = 1, b_3= 0, d_2 = 0$. 
\beq
b_2 =  1 ~, \quad  b_3= 0 ~, \quad  d_2 =0 ~. 
\eeq
\end{itemize} 
For the present application it is crucial to check that the  vertex functions 
$(\Gamma_{VV})^{abc}_{\mu \nu} (q,p)  = 
\int  d^4 x  \langle 0| T (V_\mu^a (x)  \, V_\nu^b  (0) | \pi^c (p) \rangle $
and 
$(\Gamma_{VA})^{abc}_{\mu \nu}  (q, p)= 
\int  d^4 x  \langle 0| T (V_\mu^a (x)  \, A_\nu^b  (0) | \pi^c (p) \rangle $
satisfy the correct asymptotic behavior dictated by QCD for  $q \to \infty$. 
With our ansatz  $\Gamma_{VV}$  satisfies the leading and next-to-leading power behavior 
 ($O(q^{-1})$ and $O(q^{-2})$) required by QCD. 
Concerning   $\Gamma_{AV}$,  the ansatz of Ref. ~\cite{mouss97,kn01} reproduces 
the QCD behavior to  $O(q^{-1})$ and $O(q^{-2})$. 
On the other hand,  the ansatz of Ref. ~\cite{vcetal04} has the correct $O(q^{-1})$ behavior 
but to $O(q^{-2})$ gives a result that is twice the QCD one. 
Due to these considerations,  in our analysis we will use choice 1.  above for
the $\langle V A P \rangle $ form factors and use difference in the results from choice 2. as 
an indicator of the model dependence.  We will find that the spread in result is 
minimal, showing that the convolution integral is dominated by low  and 
intermediate virtualities.  This feature is quite welcome in that it makes our results more robust.

\subsection{Results}

The matching calculation is straightforward but tedious.  It involves 
(i)  inserting  the large-$N_C$ form factors of in the convolution representation of Eq.~\ref{eq:convolution}; 
(ii)  reducing the resulting integrals to scalar Passarino-Veltman functions; 
(iii)  expanding  the full result  in powers of  $m_{\ell, \pi}/M_{V}$, up to order 
$(m/M_V)^2$;    
(iv)  finally, subtracting the ChPT$_\infty$ result from the expanded full result, thus obtaining the counterterm amplitude according to Eq.~\ref{eq:match3}. 
The details of this  calculation are reported in Appendix~\ref{sect:matchdetails}. 

Using the coefficients $b_{2,3}$ and $d_{2}$ as determined in Ref.~\cite{kn01}, and 
defining $z_A = M_A/M_V$,  we find
\bea
T_{\ell}^{e^2 p^4, CT } (\mu)  &=& T_{\ell}^{p^2}  \  \frac{\alpha}{4  \pi}  \frac{m_\ell^2}{M_V^2} 
\Bigg\{  \left[ \frac{4}{3} V_1 \, M_V^2 - \frac{7}{3 z_A^2} - \frac{11}{3} \right] 
\log \frac{M_V^2}{\mu^2}   - \frac{19}{9} V_1 \, M_V^2 
\nonumber \\
&+&  \frac{1}{18 z_A^2 \, (-1 + z_A^2)^2} \left[ -37 + 31 z_A^2 - 17 z_A^4 + 11 z_A^6 \right] 
\nonumber \\
&-&  \frac{2}{3  z_A^2 \, (-1 + z_A^2)^3} \left[ - 7 + 5  z_A^2  +   z_A^4  -    z_A^6 \right]  \log z_A
\Bigg\}  ~  . 
\label{eq:ct1}
\eea
If one uses instead the values of $b_{2,3}$ and $d_{2}$ from Ref.~\cite{vcetal04}, the counterterm amplitude is obtained by adding to  Eq.~\ref{eq:ct1} the following expression:
\beq
\delta T_\ell^{e^2 p^4, CT} = T_{\ell}^{p^2}  \  \frac{\alpha}{4 \pi}  \frac{m_\ell^2}{M_V^2} \, 
\frac{ \log z_A^2}{3 (-1 + z_A^2)}~. 
\eeq

We defer a full discussion of the implications of this result to Section~\ref{sect:pheno}. 
Here we  wish to point out that our matching  procedure captures in full the 
 "single-log" scale dependence of the counterterm as dictated by the renormalization group. 
This means that the scale dependence of Eq.~\ref{eq:ct1} cancels the bulk of the 
scale dependence from chiral loops, leading to  a very stable result.

\section{Real-photon corrections}
\label{sect:real}

\subsection{Radiative decay in ChPT}

The amplitude for the radiative decay $\pi^+ (p) \to \ell^+ (p_\ell)   \nu  (p_\nu) \gamma (q)$  
can  be written as~\cite{Bijnens:1992en}:
\bea
T_{\ell}^{\rm rad} &=& i \, 2 e G_F  F_\pi  V_{ud}^*   \, \epsilon_\mu^* (q)  \, 
 \Big(  
  B^\mu    \ - \ H^{\mu \nu} \, l_\nu 
\Big)
\\
B^\mu &=&     m_\ell \, \bar{u}_L (p_\nu)  \left[
\frac{2 p^\mu}{2  p \cdot q} - \frac{2 p_\ell^\mu + \slashed{q} \gamma^\mu}{2 p_\ell \cdot q} 
\right] v (p_\ell)
\\
H^{\mu \nu} &=& i V_1 \epsilon^{\mu \nu \alpha \beta} q_\alpha p_\beta - A_1  
\Big( q \cdot \left( p - q \right)  g^{\mu \nu}  -  \left(p - q \right)^\mu q^\nu
\Big)
\\
l_\nu &=&  \bar{u}_L (p_\nu)  \gamma_\nu  v(p_\ell)   \ , 
\eea
with the form factors $V_1$ and $A_1$  given to $O(p^4)$ in Eqs.~\ref{eq:V1} and~\ref{eq:A1}.
The part of the amplitude proportional to $B^\mu$ is referred 
to as "Inner Bremsstrahlung" (IB) component, while the 
part proportional to $H^{\mu \nu}$  is called "Structure Dependent"  (SD) component. 
IB and SD components are separately gauge invariant.  The radiative decay rate 
has a term coming from the IB amplitude squared, a term from the interference of IB and SD, 
and finally a term proportional to the SD amplitude squared. To the order we work in 
the chiral expansion, only the first two terms  have to be considered in principle,  and lead, respectively, 
to $\delta_{\ell}^{e^2 p^2}$ and $\delta_{\ell}^{e^2 p^4}$ in the expression for $R_{e/\mu}$ 
in Eq.~\ref{eq:Rmaster2}.

Introducing the dimensionless kinematical variables 
\beq
x = \frac{2 p \cdot q}{m_\pi^2}  \qquad y = \frac{2 p \cdot p_\ell}{m_\pi^2}    \ , 
\eeq 
the differential radiative decay rate is~\cite{Bijnens:1992en}
\bea
\frac{ d^2  \Gamma (\pi \to \ell \nu \gamma)}{dx \, dy} &=& \! \!  \! \! 
\frac{\alpha}{2 \pi}  \frac{\Gamma^{(0)} (\pi \to \ell \nu)}{(1 - z_\ell)^2}  
\Bigg[  \!
f_{IB} (x,y) +   m_\pi^2 \Big( V_1 \,  f_{INT}^{(V)}  (x,y)  +  A_1  \, f_{INT}^{(A)}  (x,y) \Big)  
\! \! \Bigg]
\\
f_{IB} (x,y) &=&    \frac{ 1 - y + z_\ell }{x^2  (x + y - 1 - z_\ell)} \left[ 
x^2 + 2 (1 -x)(1 - z_\ell)   - \frac{2 x z_\ell (1 - z_\ell)}{x + y - 1 - z_\ell}
\right]
\\
f_{INT}^{(V)} (x,y) &=&     \frac{x ( 1 - y + z_\ell) }{x + y - 1 - z_\ell}  \\
f_{INT}^{(A)} (x,y)  &=&    \frac{1}{x} \,  \frac{( 1 - y + z_\ell) }{x + y - 1 - z_\ell}  \left[ 2 z_\ell 
- x^2 + 2 (1-x) (1 - x - y) \right]  ~.
\eea
The total rates are obtained by integrating over the physical region 
\bea
2 \sqrt{z_\gamma}    \leq  & x & \leq  1 - z_\ell  + z_\gamma    \nonumber \\
1 - x + \frac{z_\ell}{1 - x}  \leq & y & \leq 1 + z_\ell ~. 
\eea
When integrating the IB component over the whole physical region, an infrared 
divergence arises. It must be regulated in the same way as in the virtual corrections 
(in our choice by giving an infinitesimal mass to the photon), 
and it will eventually disappear when one calculates the observable  inclusive rate. 
All integrals can be done analytically, and the results are reported in the next section.

\subsection{Results for $\delta_{\ell}^{e^2 p^2}$,  $\delta_{\ell}^{e^2 p^4}$, and 
 $\delta_{\ell}^{e^2 p^6}$}
The IB contribution to the radiative rate reads~\cite{Kinoshita:1959ha}
(recall the definition of $\delta_\ell^{e^2 p^{2n}}$ in Eq.~\ref{eq:deltae2p2n}):
\bea
\delta_\ell^{e^2 p^2} & = & \frac{\alpha}{\pi}  \, \Bigg\{
- \frac{z_\ell (10 - 7 z_\ell)}{4 ( 1 - z_\ell)^2} \log z_\ell  
+ \frac{15 - 21 z_\ell}{8 ( 1 - z_\ell)} - 
2 \frac{1 + z_\ell}{1 - z_\ell}  \, Li_2 ( 1 - z_\ell) 
\nonumber \\
&+& \left[ 
2 + \frac{1 + z_\ell}{1 - z_\ell} \log z_\ell
\right] 
\left[
\log \sqrt{z_\gamma}  - \log (1 - z_\ell)  - \frac{1}{4} \log z_\ell  + \frac{3}{4} 
\right] 
\Bigg\}~, 
\label{eq:reale2p2}
\eea
where   
\beq
Li_2 (x) =     - \int_0^{x}  \frac{ dt}{t}  \, \log (1-t)~.
\eeq
The above formula refers to the fully photon-inclusive radiative rate. If one considers 
only the radiation of soft photons with $E_\gamma^{\rm CMS} < \omega \ll m_\pi$  one finds, 
up to terms suppressed by $\omega/m_\pi$~\cite{Kuraev:1996vn}, 
\bea
\delta_\ell^{e^2 p^2} (\omega) & = & \frac{\alpha}{\pi}  \, \Bigg\{
1   - \frac{1 + z_\ell}{2(1 - z_\ell)}  \,  \log  z_\ell 
 - \frac{1 + z_\ell}{4(1 - z_\ell)}  \,  \log^2 z_\ell 
 - \frac{1 + z_\ell}{1 - z_\ell}  \, Li_2 ( 1 - z_\ell) 
\nonumber \\
&+& \left[ 
2 + \frac{1 + z_\ell}{1 - z_\ell} \log z_\ell
\right] \, \log \frac{m_\gamma}{2 \omega}
\Bigg\}~, 
\label{eq:reale2p2soft}
\eea

The interference between IB and SD amplitude (parameterized in terms of the 
form factors $V_1$ and $A_1$)   reads: 
\bea
\delta_\ell^{e^2 p^4} & = & 
 \frac{\alpha}{2 \pi} \, \frac{m_\pi^2}{(1 - z_\ell)^2} \,  
\Bigg\{
V_1  \, 
\left[
- \frac{17}{18}  + \frac{z_\ell}{2}  + \frac{z_\ell^2}{2}  - \frac{z_\ell^3}{18}  -
\frac{1}{3} \log z_\ell  -  z_\ell   \log z_\ell     \right]
\nonumber \\
&+&  A_1  \, 
\left[
 \frac{7}{9}  - 2 z_\ell + z_\ell^2   + \frac{2\, z_\ell^3}{9}  +
\frac{1}{3} \log z_\ell  -  z_\ell^2     \log z_\ell  
 \right]
\Bigg\}
\label{eq:reale2p4v1}~.
\eea
Classifying the various terms according to their behavior with the lepton mass, 
one obtains:
\bea
\delta_\ell^{e^2 p^4} & = & 
\frac{\alpha}{2 \pi}  \left( \frac{7}{9} A_1 - \frac{17}{18} V_1 \right) \,  m_\pi^2
+ 
\frac{\alpha}{2 \pi}  \left( A_1 - V_1 \right) \,  \frac{m_\pi^2}{3} \log z_\ell  
\nonumber \\
&+&  
\frac{\alpha}{2 \pi}  \left( - \frac{4}{9} A_1 - \frac{25}{18} V_1 \right) \,  m_\ell^2
+ 
\frac{\alpha}{2 \pi}  \left(2  A_1 - 5  V_1 \right) \,  \frac{m_\ell^2}{3} \log z_\ell  
\nonumber \\   
&+&  \frac{\alpha}{2 \pi} \, m_\ell^2  \,   \frac{z_\ell}{(1 - z_\ell)^2} 
\bigg\{
- \frac{2}{3} A_1  \, \Big( 1 - z_\ell  + z_\ell \log z_\ell \Big)   \nonumber \\
&-&  \frac{1}{3} V_1  \, 
\Big[
4 ( 1 - z_\ell)   + ( 9 - 5 z_\ell )  \log z_\ell 
 \Big]
\bigg\}~.
\label{eq:reale2p4}
\eea

Finally, we report here  the purely SD contribution to the radiative 
rate, which is down by one order in the chiral expansion  but does not suffer from 
helicity suppression.  We find:
\bea
\delta_\ell^{e^2 p^6}  &=&  
\frac{\alpha}{8 \pi}  \, m_\pi^4 \left(V_1^2 + A_1^2 \right)  \ 
\Bigg[ 
\frac{1}{30 \, z_\ell} - \frac{11}{60} 
\nonumber \\
&+& 
\frac{z_\ell}{20 (1 - z_\ell)^2}  
\left( 
12 - 3 z_\ell - 10 z_\ell^2 + z_\ell^3 + 20\,  z_\ell \log z_\ell 
\right)
\Bigg]~.
\eea

\section{Phenomenology of $R_{e/\mu}$}
\label{sect:pheno}

We now put together all the results obtained so far.  
The starting point of our phenomenological analysis of  $R_{e/\mu}^{(\pi,K)}$ is
Eq.~\ref{eq:Rmaster1},  which organizes the electroweak corrections to the leading order 
result of Eq.~\ref{eq:R0} according to the chiral expansion. 
Incorporating the effects of leading higher order logs~\cite{MS93}
 of the form $\alpha^n \log^n (m_\mu/m_e)$
through the correction $\Delta_{LL}$, we can write Eq.~\ref{eq:Rmaster1} as:
\beq
R_{e/\mu}^{(P)} = R_{e/\mu}^{(0),(P)}   
\, \Big[   1 + 
\Delta_{e^2 p^2}^{(P)} +   \Delta_{e^2 p^4}^{(P)}  + \Delta_{e^2 p^6}^{(P)}   +  ...  
\Big]  \Big[ 1 + \Delta_{LL}  \Big]\ \ \ \  
\label{eq:Rmaster3}
\eeq

The leading electromagnetic correction in ChPT 
 corresponds to  the point-like approximation for pion and  
 kaon~\cite{MS93,Knecht:1999ag,Kinoshita:1959ha}: 
\bea
\Delta^{(P)}_{e^2 p^2} &= & \frac{\alpha}{\pi}  \Big[ F( \frac{m_e^2}{m_P^2}) - F(\frac{m_\mu^2}{m_P^2})
 \Big]   \\
F(z)   &=&  \frac{3}{2} \log z  + \frac{13 - 19 z}{8 ( 1 - z)} - 
\frac{8 - 5 z}{4 (1 - z)^2}  \, z   \log z
- \left(  2 + \frac{1 + z}{1 - z} \log z  \right)  \log ( 1 - z) 
\nonumber \\
&-& 2 \frac{1 + z}{1 - z}  \, Li_2  ( 1 - z)     ~.
\eea

The structure dependent effects are all contained in 
$\Delta_{e^2 p^4}$ and higher order terms,  which are the main 
subject of this work. 
Neglecting terms of order $(m_e/m_\rho)^2$,  the most general parameterization of the 
NLO  ChPT contribution can be written in the form 
\beq
 \Delta_{e^2 p^4}^{(P)} = \frac{\alpha}{\pi} \frac{m_\mu^2}{m_\rho^2} 
\left(c_2^{(P)}  \, \log \frac{m_\rho^2}{m_\mu^2}  
+  c_3^{(P)} 
+ c_4^{(P)} (m_\mu/m_P) \right) + 
\frac{\alpha}{\pi}  \frac{m_P^2}{m_\rho^2} \,  \tilde{c}_{2}^{(P)}  \, \log \frac{m_\mu^2}{m_e^2} ~ , 
\label{eq:dele2p4}
\eeq
which highlights the dependence on lepton masses. 
The dimensionless constants  $c_{2,3}^{(P)}$  do not  depend on the lepton mass 
but depend logarithmically on hadronic masses, while  $c_4^{(P)} (m_\mu/m_P) \to 0$ 
as $m_\mu \to 0$. 
(Note that  our $c_{2,3}^{(\pi)}$  do not coincide with 
$C_{2,3}$ of Ref.~\cite{MS93}, because their $C_{3}$
is not constrained to be $m_\ell$-independent.)

Finally, let us note that the  results for $c_{2,3,4}^{(P)}$ and $\tilde{c}_{2}^{(P)}$   depend on the definition of the inclusive 
rate  $\Gamma (P \to \ell \bar{\nu}_\ell [\gamma])$.  
The radiative  amplitude is the sum of the  inner bremsstrahlung ($T_{IB}$) component 
of $O(e p)$ and a structure 
dependent ($T_{SD}$) component of $O(e p^3)$~\cite{Bijnens:1992en}.   
The experimental  definition of $R_{e/\mu}^{(\pi)}$ is fully inclusive on the radiative mode, so 
that $\Delta_{e^2 p^4}^{(\pi)}$ receives a contribution  from the interference
of $T_{IB}$ and $T_{SD}$.  
Moreover, in this case 
one also has to include the effect of $\Delta_{e^2 p^6}^{(\pi)} \propto |T_{SD}|^2$, 
that is formally of $O(e^2 p^6)$, but is not helicity suppressed 
and behaves as $\Delta_{e^2 p^6} \sim   \alpha/\pi  \,  (m_P/M_V)^4  \, (m_P/m_e)^2$.
On the other hand, the usual experimental definition of  $R_{e/\mu}^{(K)}$ 
is not fully inclusive on the radiative mode.  It 
corresponds to including the effect of 
$T_{IB}$ in $\Delta_{e^2 p^2}^{(K)}$ (dominated by soft photons) and excluding altogether 
the effect of $T_{SD}$: consequently $c_n^{(\pi)} \neq c_n^{(K)}$. 

\subsection{Results for $R_{e/\mu}^{(\pi)}$}

Recalling the definitions  $\bar{L}_9 \equiv (4 \pi)^2  L_{9}^r (\mu)$, 
$\ell_P \equiv \log (m_P^2/\mu^2)$  ($\mu$ is the chiral renormalization scale),   
$\gamma \equiv  A_1(0,0) / V_1 (0,0)$,  $z_\ell \equiv (m_\ell/m_\pi)^2$, we find:   
\bea
c_2^{(\pi)} \!\!\!&=& \!\!\!
\frac{2}{3} \, m_\rho^2  \, \langle r^2 \rangle_{V}^{(\pi)}
+   3 \, \left( 1 - \gamma \right)   \, \frac{m_\rho^2}{(4 \pi F)^2}    \qquad \qquad \qquad 
\tilde{c}_2^{(\pi)}  \,  =  \, 0  
\\
c_3^{(\pi)}\!\!\! &=& \!\!\!-  \frac{m_\rho^2}{(4 \pi F)^2}  
\Bigg[  \frac{31}{24} - \gamma + 4 \, \bar{L}_9   + 
\left(\frac{23}{36}  - 2 \, \bar{L}_9  + \frac{1}{12} \ell_K \right) \ell_\pi  + \frac{5}{12} \ell_\pi^2 
 + \frac{5}{18} \ell_K +  \frac{1}{8} \ell_K^2  
\nonumber \\
\!\!\!&+& \!\!\!\left( \frac{5}{3} - \frac{2}{3} \gamma \right) \, \log \frac{m_\rho^2}{m_\pi^2} 
+ \left( 2 + 2 \,  \kappa^{(\pi)} - \frac{7}{3} \gamma  \right) \, \log \frac{m_\rho^2}{\mu^2} 
+  K^{(\pi)} (0)   \Bigg]  \ + \ c_{3}^{CT} (\mu) 
\label{eq:C3pi}
\\
c_4^{(\pi)} (m_\ell) \!\!\!&=& \!\!\!
-  \frac{m_\rho^2}{(4 \pi F)^2}   \left\{ 
\frac{z_\ell}{ 3  (1 - z_\ell)^2} 
\left[    \Big( 4 ( 1 - z_\ell)   + ( 9 - 5 z_\ell )  \log z_\ell  \Big)
+ 2 \,   \gamma  \,  \Big( 1 - z_\ell  + z_\ell \log z_\ell \Big)    \right] 
\right. 
\nonumber \\
\!\!\!&+& \!\!\!\left.  \left( \kappa^{(\pi)}  + \frac{1}{3} \right) \frac{  \, z_\ell}{2 ( 1 - z_\ell)} \, \log z_\ell   
+ K^{(\pi)}  (m_\ell) - K^{(\pi)} (0)   \right\} 
\label{eq:C4pi}
\eea
where $\kappa^{(\pi)}$ is related to the $O(p^4)$ pion charge radius by:  
\beq
\kappa^{(\pi)} \equiv  4 \, \bar{L}_9 - \frac{1}{6} \ell_K - \frac{1}{3} \ell_\pi - \frac{1}{2} 
=   \frac{(4 \pi F_\pi)^2} {3}  \, \langle r^2 \rangle_{V}^{(\pi)} ~.
\label{eq:klogg}
\eeq
In the above equations we have used the definition:
\beq
K^{(\pi)} (m_\ell) =  \frac{1}{2}  \left[ f_1 (z_\ell) + \frac{1}{2} f_1 (\tilde{z}_\ell) 
+ f_2 (z_\ell) + f_3 (\tilde{z}_\ell, \tilde{z}_\pi) 
-\frac{8}{9}  \log \frac{m_\rho^2}{m_\pi^2} 
-\frac{4}{9}  \log \frac{m_\rho^2}{m_K^2} 
\right] ~.
\eeq
The function $K^{(\pi)} (m_\ell)$ does not contain any large logarithms 
($K^{(\pi)} (m_\mu) = -0.025$ and $K^{(\pi)} (0) = 0.085$)
and gives a small  fractional contribution to $c_{3,4}^{(\pi)}$.  

Full numerical values of $c_{2,3,4}^{(\pi)}$ and $\tilde{c}_2^{(\pi)}$ are reported in Table~\ref{tab:tab1}, with 
uncertainties due to matching  procedure 
 and input parameters ($L_9$ and $\gamma$~\cite{pocanic}). We now discuss the results obtained and make contact with the previous literature. 
\begin{itemize}
\item   We find $\tilde{c}_2^{(\pi)} = 0$ in accordance to a theorem by Marciano and 
Sirlin~\cite{Marciano:1976jc}.  This result arises from an exact cancellation of    
virtual  photon contributions proportional to $V_1$ and $A_1$  and real photon 
contribution due to the interference of SD and IB amplitudes. 
The cancellation occurs only when the fully inclusive rate is considered.

\item  The coefficient $c_2^{(\pi)}$   is a parameter-free prediction of ChPT to this order. 
It involves the $O(p^4)$ LECs  $L_9$ and $L_{10}$, related to the pion charge radius and the 
ratio of axial-to-vector form factors $\gamma$ measurable in the radiative pion and kaon decay.

\begin{table}[t!]
\begin{center}
\begin{tabular}{|c|c|c|}
\hline
  & $(P=\pi)$  & $(P=K)$    \\[5pt]
\hline
 $\tilde{c}_2^{(P)}$  &   0  &  $ (7.84 \pm 0.07_\gamma) \times 10^{-2}  $ \\
 $c_2^{(P)}$  & $5.2 \pm 0.4_{L_9} \pm 0.01_\gamma$  &  $4.3 \pm 0.4_{L_9} \pm 0.01_\gamma $ \\
 $c_3^{(P)}$   
&   
$ -10.5 \pm 2.3_{\rm m } \pm 0.53_{L_9} 
$
& 
$ -4.73 \pm 2.3_{ \rm m} \pm 0.28_{L_9}$ 
\\
 $c_4^{(P)} (m_\mu) $  &
$1.69 \pm 0.07_{L_9} $
&  
$ 0.22 \pm 0.01_{L_9} $ 
\\
\hline
\end{tabular}
\end{center}
\caption{Numerical values for $c_n^{(P)}$ of Eq.~\ref{eq:dele2p4}, for $P=\pi,K$.
The uncertainties correspond to the input values  $L_9^r (\mu=m_\rho) = 
(6.9 \pm 0.7) \times 10^{-3} $, 
$\gamma= 0.465 \pm 0.005$~\cite{pocanic}, and to the matching procedure (${\rm m}$), 
affecting only $c_3^{(P)}$.}
\label{tab:tab1}
\end{table}

\item The coefficient $c_3^{(\pi)}$  receives  a predictable contribution from loops in the ChPT framework,  as well as  a local contribution that cannot be predicted in the purely EFT approach, 
denoted by $c_3^{CT}(\mu)$.  Both contributions are renormalization-scale dependent, while the 
sum is not.      
$c_{3}^{CT}(\mu)$   is related to the low energy coupling $r_{CT}(\mu)$ introduced in 
Eq.~\ref{eq:rCT} by  $r_{CT} (\mu) = - 2   (4 \pi F)^2/m_\rho^2  \ c_3^{CT}(\mu) $.
Our matching procedure gives for the counterterm
($z_A \equiv  (M_A/M_V)$ and taking $M_V=m_\rho$):
\bea
c_3^{CT} (\mu) \! &=& \!
-  \frac{19 \, m_\rho^2}{9 (4 \pi F)^2}   +  \left(  \!
\frac{ 4 \, m_\rho^2}{3 (4 \pi F)^2}  +  \frac{7 + 11 z_A^2}{6      z_A^2}  \! 
\right)   \log\frac{m_\rho ^2 }{\mu^2}  
\nonumber \\
 &+& \frac{ 37 - 31 z_A^2   + 17 z_A^4 - 11 z_A^6}{ 36 z_A^2  ( 1 - z_A^2)^2}
 \nonumber \\
&-&  \frac{ 7 - 5 z_A^2   - z_A^4  + z_A^6}{ 3 z_A^2  ( - 1 + z_A^2)^3} \, \log z_A ~.
\label{eq:CT}
\eea
Numerically,  using  $z_A = \sqrt{2}$~\cite{Ecker:1989yg}, we find $c_3^{CT} (m_\rho) = -1.61$,   
implying that the counterterm induces a sub-leading correction to $c_3$ (see Table~\ref{tab:tab1}). 
The model dependence due to different choices of the hadronic ansatz (Ref.~\cite{kn01} vs 
Ref.~\cite{vcetal04})  is negligible, being  $\Delta c_3^{CT} = 0.12$. 
The scale dependence of 
$c_{3}^{CT}(\mu)$ 
partially cancels the scale dependence 
of the  chiral loops (our procedure captures all the "single-log" scale dependence). 
Taking a very conservative attitude we assign to $c_3$  an uncertainty equal to 
$100\%$  of the local contribution ($| \Delta c_3 | \sim 1.6$)  plus the effect of residual 
renormalization scale dependence, obtained by varying the scale $\mu$ in the range 
$0.5 \to 1$ GeV   ($| \Delta c_3|  \sim 0.7$), leading to 
$\Delta c_3^{(\pi,K)} = \pm 2.3$. 

\item  Finally,  the coefficient $c_4^{(\pi)}$  can be calculated in terms of the LEC $L_9$ 
and the lepton and meson 
masses and decay constants. 
Not surprisingly we find that this effect is only marginally important.

\end{itemize}

As a check on our calculation, we have verified that if we neglect $c_{3}^{CT}$ and pure two-loop effects,  and if we use $L_9 = F^2/(2 M_V^2)$  (vector meson dominance), 
our results for $c_{2,3,4}^{(\pi)}$ are   
fully consistent with previous analyses of the leading 
structure dependent corrections based on current algebra~\cite{MS93,terentev}.    
Moreover, our numerical value of  
$\Delta_{e^2 p^4}^{(\pi)}$ reported in  Table~\ref{tab:tab2} is very close to the corresponding 
result  in Ref.~\cite{MS93}, namely 
$\Delta_{e^2 p^4}^{(\pi)} =  (0.054 \pm 0.044) \times 10^{-2}$~\cite{MS93} 
versus 
$\Delta_{e^2 p^4}^{(\pi)} =  (0.053 \pm 0.011) \times 10^{-2}$~(this work). 
Therefore, as far as $R_{e/\mu}^{(\pi)}$ is concerned, 
the net effect of our calculation is a reduction of the uncertainty by a factor of four.

\subsection{Results for $R_{e/\mu}^{(K)}$}
In the case of $K$ decays we find:

\bea
c_2^{(K)}\!\!\! &=&\!\!\!
\frac{2}{3} \, m_\rho^2  \, \langle r^2 \rangle_{V}^{(K)}
+ \frac{4}{3} \,  \left( 1 - \frac{7}{4} 
\gamma \right)\, \frac{m_\rho^2}{(4 \pi F)^2}   
\\
\tilde{c}_2^{(K)} \!\!\! &  =  & \!\!\! \frac{1}{3} \left(1 - \gamma \right)  \,
 \frac{m_\rho^2}{(4 \pi F)^2}   
 \\
 c_3^{(K)} \!\!\!&=& \!\!\!-  \frac{m_\rho^2}{(4 \pi F)^2}
\Bigg[ - \frac{7}{72} - \frac{13}{9} \gamma +  4 \, \bar{L}_9    + 
\left(\frac{23}{36}  - 2  \bar{L}_9  + \frac{1}{12} \ell_\pi \! \right) \ell_K  + \frac{5}{12} \ell_K^2 
 + \frac{5}{18} \ell_\pi +  \frac{1}{8} \ell_\pi^2   
 \nonumber \\
\!\!\!&+&\!\!\!
\left( 2 + 2 \,  \kappa^{(K)} - \frac{7}{3} \gamma  \right) \, \log \frac{m_\rho^2}{\mu^2} 
+  K^{(K)}(0)   \Bigg]  \ + \ c_{3}^{CT} (\mu) 
\\
c_4^{(K)} (m_\ell) \!\!\!&=& \!\!\!
-  \frac{m_\rho^2}{(4 \pi F)^2}   \left\{ 
   \left( \kappa^{(K)} + \frac{1}{3} \right) \frac{  \, \tilde{z}_\ell}{2 ( 1 - \tilde{z}_\ell)} \, \log 
 \tilde{z}_\ell    +  K^{(K)} (m_\ell) - K^{(K)} (0)   \right\} ~, 
\eea
where $\langle r^2 \rangle_{V}^{(K)}$
is the $O(p^4)$ kaon charge radius and 
\beq
\kappa^{(K)} \equiv  4 \, \bar{L}_9 - \frac{1}{6} \ell_\pi - \frac{1}{3} \ell_K - \frac{1}{2} 
 =   \frac{(4 \pi F)^2} {3}  \, \langle r^2 \rangle_{V}^{(K)} ~.
\label{eq:klog2}
\eeq
Moreover the  function $K^{(K)} (m_\ell)$ is given by:
\beq
K^{(K)} (m_\ell) =  \frac{1}{2}  \left[ f_1 (\tilde{z}_\ell) + \frac{1}{2} f_1 (z_\ell) 
+ f_2 (\tilde{z}_\ell) + f_3 (z_\ell, 1/\tilde{z}_\pi) 
-\frac{8}{9}  \log \frac{m_\rho^2}{m_K^2} 
-\frac{4}{9}  \log \frac{m_\rho^2}{m_\pi^2} 
\right] ~.
\eeq
As in the pion case, the function $K^{(K)} (m_\ell)$ does not contain any large logarithms 
($K^{(K)} (m_\mu) = 0.93$ and $K^{(\pi)} (0) = 1.05$)
and gives a small  fractional contribution to $c_{3,4}^{K}$.  

Note that apart from missing contributions from the SD radiation, the 
$c_{2,3,4}^{(K)}$ and $\tilde{c}_2^{(K)}$ are obtained from the $c_{2,3,4}^{(\pi)}$ and  $\tilde{c}_2^{(\pi)}$ by interchanging 
$m_\pi$ with $m_K$ everywhere 
(the underlying reason is given in Sect.~\ref{sect:K}).  
The numerical values of $c_{2,3,4}^{(K)}$ and $\tilde{c}_2^{(K)}$ are reported in Table~\ref{tab:tab1}.

\subsection{Resumming long distance logarithms} 

At  the level of uncertainty considered,  one needs to include  higher order 
long distance corrections~\cite{MS93}, generalizing the leading contribution 
$\Delta_{e^2 p^2} \sim - 3 \alpha/\pi \log m_\mu/m_e \sim - 3.7 \%$.  
The leading logarithms  can be summed via the  renormalization group 
and their effect amounts to  multiplying $R_{e/\mu}^{(P)}$ by $1 + \Delta_{LL}$, 
with~\cite{MS93}
\beq
1 + \Delta_{LL} = 
\displaystyle\frac{\left(1 - \frac{2}{3} \frac{\alpha}{\pi} \log \frac{m_\mu}{m_e} \right)^{9/2}}{1 -
 \frac{3 \alpha}{\pi} \log \frac{m_\mu}{m_e}} = 
1.00055~.
\eeq

\subsection{Discussion}

\begin{table}[t!]
\begin{center}
\begin{tabular}{|c|c|c|}
\hline
 & $(P=\pi)$  & $(P=K)$    \\[5pt]
\hline
 $\Delta_{e^2 p^2}^{(P)} \ \, (\%)  $   &   $-3.929 $ &  $ -3.786 $ \\
$\Delta_{e^2 p^4}^{(P)}  \ \, (\%) $ 
 & $0.053 \pm 0.011$  &  $0.135 \pm 0.011$ \\
$\Delta_{e^2 p^6}^{(P)} \ \,  (\%)  $ 
&   
$ 0.073$
& 
\\
$\Delta_{LL} \ \ (\%) $   &
$ 0.055 $ 
&  
$ 0.055 $ 
\\
\hline
\end{tabular}
\end{center}
\caption{Numerical summary of various electroweak corrections to $R_{e/\mu}^{(\pi,K)}$. 
The uncertainty in $\Delta_{e^2 p^4}$ corresponds to the matching procedure. }
\label{tab:tab2}
\end{table}

In Table~\ref{tab:tab2} we summarize the various electroweak corrections to  $R_{e/\mu}^{(\pi,K)}$.    
Applying these  we arrive to our final results:
\bea
R_{e/\mu}^{(\pi)} &=& (1.2352 \pm 0.0001 ) \times 10^{-4} 
\label{eq:Rpf} \\ 
R_{e/\mu}^{(K)} &=& ( 2.477 \pm 0.001 ) \times 10^{-5} ~ . 
\label{eq:RKf}
\eea
The uncertainty we quote for $R_{e/\mu}^{(\pi)}$  is entirely induced by our matching procedure. 
However, in the case of $R_{e/\mu}^{(K)}$ we have inflated the nominal uncertainty  
arising from matching  by a factor of four,  to account for higher order chiral corrections, that  
are expected  to scale as  $\Delta_{e^2 p^4} \times m_K^2/(4 \pi F)^2$.   
 
Our results have 
to be compared with the ones of Refs.~\cite{MS93} and \cite{Fink96}, which we report 
in Table~\ref{tab:tab3}. 
While  $R_{e/\mu}^{(\pi)}$ is  in good agreement with both previous results,  
there is a discrepancy  in $R_{e/\mu}^{(K)}$ that goes well outside the 
estimated theoretical uncertainties. 
We have traced back  this difference  to  two problematic aspects of  Ref.~\cite{Fink96}. 
(i) The leading log correction $\Delta_{LL}$ is included  with the wrong sign: this accounts for half  
of the discrepancy.   
(ii) The remaining effect is due to the difference in the NLO virtual correction, for which  Finkemeier 
finds $\Delta_{e^2 p^4}^{(K)}  =  0.058 \%$. 
We have serious doubts on the reliability of this number  because the hadronic form factors 
modeled in Ref.~\cite{Fink96}  do not satisfy the correct QCD short-distance behavior. 
At high momentum they  fall off faster than the QCD requirement, thus 
leading to a smaller value of  $\Delta_{e^2 p^4}^{(K)}$ compared to our work. 
\begin{table}[t!]
\begin{center}
\begin{tabular}{|c|c|c|}
\hline
 & $10^4 \cdot R_{e/\mu}^{(\pi)}$  & $ 10^5 \cdot  R_{e/\mu}^{(K)}$    \\[5pt]
\hline
 This work   &    $1.2352 \pm 0.0001$ & 
$ 2.477 \pm 0.001 $ \\
Ref.~\cite{MS93}   & 
$1.2352 \pm 0.0005$ 
 & 
 \\
Ref.~\cite{Fink96}
&   
$1.2354 \pm 0.0002$ 
&  
$ 2.472 \pm 0.001 $ 
\\
\hline
\end{tabular}
\end{center}
\caption{Comparison of our result with the most  recent predictions of  $R_{e/\mu}^{(\pi,K)}$.}
\label{tab:tab3}
\end{table}

\section{The individual $\pi (K) \to \ell \bar{\nu}_\ell$ modes}
\label{sect:update}

The approach followed in this work is  designed to obtain the ratio 
of $\pi (K) \to e \bar{\nu}_e$ and $\pi (K) \to \mu \bar{\nu}_\mu$ 
decay rates, because we have neglected all the Feynman diagrams 
in which the photon does not connect to the charged lepton.  
Including these diagrams in ChPT would generate new finite parts and  UV divergences,  
and the corresponding local couplings would have to be evaluated 
within the $1/N_C$ expansion described earlier. 
We leave this task  for  possible future work.

However, despite the fact that we have not performed a full $O(e^2 p^4)$ calculation of 
$\pi (K) \to \ell \bar{\nu}_\ell$,  our results can still be used to update the 
theoretical analysis of these individual decay modes. 
Here we closely follow the analysis of Ref.~\cite{MS93}. 
Including all known short- and long-distance electroweak 
corrections, and parameterizing the hadronic effects in terms of a few dimensionless 
coefficients,    the inclusive 
$P \to \ell \bar{\nu}_\ell [\gamma] $ decay rate $\Gamma_{P_{\ell 2 [\gamma]} }$ 
can be written as:
\bea
\Gamma_{P_{\ell 2 [\gamma]}}
\!\!\!\!
&=& 
\Gamma^{(0)} \times 
\Bigg\{ 1 + \frac{2 \, \alpha}{\pi}  \log \frac{m_Z}{m_\rho} \Bigg\}
\times 
\Bigg\{
1 + \frac{\alpha}{\pi}  \,  F (m_\ell^2/m_P^2) 
\Bigg\}
\times
\Bigg\{  1 - \frac{\alpha}{\pi}  
\Bigg[
\frac{3}{2}  \log \frac{m_\rho}{m_P}  
\nonumber \\
& + & 
 c_1^{(P)}  + 
\frac{m_\ell^2}{m_\rho^2}  
\left(c_2^{(P)}  \, \log \frac{m_\rho^2}{m_\ell^2}  
+  c_3^{(P)} 
+ c_4^{(P)} (m_\ell/m_P) \right) 
-  \frac{m_P^2}{m_\rho^2} \,  \tilde{c}_{2}^{(P)}  \, \log \frac{m_\rho^2}{m_\ell^2} 
\Bigg]
\Bigg\} ~,
\label{eq:indrate}
\eea
where $\Gamma^{(0)}$  is the rate in absence of radiative corrections (see Eq.~\ref{eq:gamma0}), 
the first bracketed term is the universal short distance electroweak correction, 
the second bracketed term is the universal long distance correction (point-like meson), 
and the third bracketed term parameterizes the effects of hadronic structure. 
The function $F(z)$ and the constants  $c_{2,3,4}^{(P)}$ (and $\tilde{c}_2^{(P)}$) 
already appear in $R_{e/\mu}^{(P)}$ and their expressions and numerical 
values have been reported in the previous section. 
The only additional ingredient needed to predict
 the individual rates  $\Gamma_{P_{\ell 2 [\gamma]} }$ 
is the structure-dependent coefficient $c_1^{(P)}$, which does not depend on the lepton 
mass and starts at $O(e^2 p^2)$ in ChPT. 
The explicit form  (for both $P= \pi, K$) is given in~\cite{Knecht:1999ag} (Eqs. 5.11 and 5.14)
and it depends on a combination of EM LECs of $O(e^2 p^2)$. 
These have been recently estimated in Ref.~\cite{dm2005} in the same large-$N_C$ 
framework adopted here, with the final result:
\bea
c_1^{(\pi)} &=&   -2.56  \pm 0.5 \\
c_1^{(K)} &=&   -1.98    \pm 0.5 ~ .
\eea
So at the moment  all the structure dependent coefficients $c_n^{(p)}$ are known 
to leading order in their expansion within the chiral effective theory  
(which is $O(e^2p^2)$ for $c_1^{(P)}$ and $O(e^2p^4)$ for the other coefficients) .
The  $O(e^2 p^4)$  contribution to $c_1^{(P)}$  has not yet been calculated 
(this could be done by employing the techniques presented in this paper). 
On the basis of power counting we expect  $ c_1^{(P)} |_{e^2 p^4} \lsim 0.5$, which is 
consistent with  the uncertainty assigned to $c_1^{(P)}$~\cite{dm2005}.

Finally, we discuss  here  a quantity of interest in the experimental 
analysis of $K_{e2}/K_{\mu 2}$, namely the $K \to \ell \nu$ rate with inclusion of 
only soft photons ($\omega \ll m_K$): 
\beq
\Gamma_{K_{\ell2[\gamma]}} (\omega) \equiv
\Gamma (K \to \ell \bar{\nu}_\ell) + \Gamma (K \to \ell \bar{\nu}_\ell \gamma) 
\Big|_{E_\gamma^{\rm CMS} < \omega} ~.
\eeq
Using our results on the emission of soft photons (Eq.~\ref{eq:reale2p2soft}),  
it is simple to show that 
$\Gamma_{K_{\ell2[\gamma]}} (\omega)$  is given by Eq.~\ref{eq:indrate} 
provided one replaces $F(z) \to F^{\rm soft} (z ; \omega)$, with ($z=m_\ell^2/m_K^2$):
\bea
F^{\rm soft} (z; \omega) &=&  - \frac{3}{4} + \frac{3}{4} \log z - \frac{2 \, z}{1 - z} \log z 
- \frac{1+z}{1 - z} Li_2 (1 - z)  
\nonumber \\
&-& 
   \left[ 
2 + \frac{1 + z}{1 - z} \log z
\right] \,   \log \frac{2 \omega}{m_K}  ~.
\eea

\section{Conclusions} 
\label{sect:conclusions}
In conclusion,  by performing the first  ChPT calculation to $O(e^2 p^4)$ 
and a matching calculation of the relevant low energy  coupling, 
we have improved the reliability of both the central value and the uncertainty 
of the  ratios $R_{e/\mu}^{(\pi,K)}$.  
Our final result for $R_{e/\mu}^{(\pi)}$ is consistent with the previous literature, 
while we find a discrepancy in $R_{e/\mu}^{(K)}$,  which we have traced back to 
inconsistencies in the analysis of Ref.~\cite{Fink96}. 
Our results provide a clean basis to detect or constrain non-standard physics 
in these modes by comparison with upcoming experimental measurements.

As a byproduct of  our main analysis, we also updated the expressions for the radiative corrections 
to the individual  $\pi (K) \to \ell \bar{\nu}_\ell$ modes,  which can be used to extract from 
experiment  the combinations  $F_\pi \, V_{ud}$ and  $F_K\, V_{us}$. 

Finally,  it is worth mentioning that  the ideas and techniques discussed in  
this article can be applied (i) to perform a full $O(e^2 p^4)$ analysis  of the individual 
 $\pi (K) \to \ell \bar{\nu}_\ell$ modes; 
 (ii) to deal with other  processes that involve one pseudo-scalar meson and a lepton pair, 
such as  $\tau \to K \nu_\tau  [\gamma]$.  In this case  
chiral effective theory techniques are not adequate,   but 
the  calculation based on the 
large-$N_C$ representation for the $\langle VAP \rangle$ and $\langle VVP \rangle$
remains adequate. 

{\bf Acknowledgments}  --  We wish to thank M.~ Ramsey-Musolf for collaboration at  an early 
stage of this work,  D.~Pocanic and M.~Bychkov for correspondence on the 
experimental input on $\gamma$, and W.~Marciano and A.~Sirlin for  cross-checks on parts 
of our calculation.  
V.C. thanks Doug Bryman,  
Terry Goldman,  Evgueni Goudzovski,  Ben Grinstein,   
Gino Isidori, Marc Knecht  and Helmut Neufeld  for useful discussions. 
I.R. thanks the FPU program (Spanish MEC) for supporting his visit to Caltech, where this project started. 
He also thanks people from Caltech for their  hospitality during his stay. 
This work has been supported in part by the EU MRTN-CT-2006-035482
(FLAVIAnet), by MEC (Spain) under grant FPA2004-00996 and by Generalitat
Valenciana under grant GVACOMP2007-156.

\appendix 

\section{Two-loop integrals}
\label{sect:technicalities}
\subsection{Procedure}

In order to calculate the  genuine two-loop integrals listed 
in Sect.~\ref{sect:basic} above, 
we use the d-dimensional  dispersive representation of the function
$\bar{J}^{aa}(q^2)$~\cite{Bijnens:1997vq,Gasser:1998qt}, 
which is easily derived from Eq.~\ref{eq:Jdisp1}. 
Re-expressing all dimensionful parameters in units of $m_a$, one obtains:
\bea
\label{eq:Jdisp2}
\bar{J}^{aa}(q^2) &=&  -  m_a^{2w} \bar{q}^2 
 \int_{4}^{\infty} \, \frac{\left[ d s \right]}{s} \,  \frac{1}{(\bar{q}^2 - s)} \\
\left[ d s \right] & = &  \frac{d s}{(4 \pi)^{2+w}} \, \frac{\Gamma \left(\frac{3}{2} \right)}{\Gamma 
\left(\frac{3}{2} + w\right)} \, \left(\frac{s}{4} - 1 \right)^w \left( 1- \frac{4}{s} 
\right)^{1/2}  \ , 
\eea
where $\bar{q} = q/m_a$ and $s$ are dimensionless variables. 
Upon inserting the representation of Eq.~\ref{eq:Jdisp2} in the expression for $I_{n}^{aa}$ 
one immediately sees that the calculation is  naturally separated in two steps:
(i)  a one-loop diagram involving one propagator of  mass $s$; 
(ii) integration of the result over the variable $s$, with measure given by  $[ds]/s$. 

In order to exemplify the procedure,  we report here the calculation of $I_{1}^{(\ell) \pi \pi}$. 
The other integrals can be  worked out with similar techniques. 
We have found extremely useful the results of Ref.~\cite{Gasser:1998qt}. 
The case considered in that paper is slightly easier, because they only have 
one mass scale in the loops ($m_\pi$), while we have two. 

Inserting the representation of  Eq.~\ref{eq:Jdisp2}  in the definition of $I_{1}^{(\ell) \pi \pi}$,  
and re-expressing {\it all} momentum variables in units of $m_\pi$, one arrives at 
(recall $d = 4 + 2 w$):
\beq
I_{1}^{(\ell) \pi \pi} = - m_\pi^{4 w}  \, \int_{4}^{\infty} \frac{[ds]}{s} \, 
\int \frac{d^d q}{(2 \pi)^d} \, \frac{1}{q^2 - s} \frac{1}{\left[ (q - p_\ell)^2 - z_\ell   \right]}
\eeq
Here $q$ and $p_\ell$  are  dimensionless momentum variables 
(to avoid clutter we are not using the $\bar{q},\bar{p_\ell}$ 
notation)  and $z_\ell = (m_\ell/m_\pi)^2$. 
Combining the two denominators with the usual trick one gets:
\bea
I_{1}^{(\ell) \pi \pi}  &=&  - m_\pi^{4 w}  \, \int_{4}^{\infty} \frac{[ds]}{s} \, 
\int_{0}^{1} d x  \int \frac{d^d q}{(2 \pi)^d} \,
 \frac{1}{\left[(q - x p_\ell)^2 - z(x,s) \right]^2} 
\nn
&=& 
 - i \, m_\pi^{4 w}  \, \int_{4}^{\infty} \frac{[ds]}{s} \, 
\int_{0}^{1} d x  \, F_2 [z(x,s)]     
\label{eq:I1a} 
\eea
where $z(x,s) = z_\ell  \, x^2 + s (1 - x)$ 
and~ \cite{Gasser:1998qt}
\beq
i (-1)^n F_n [z] =  \int \frac{d^d q}{(2 \pi)^d} \,
 \frac{1}{\left[q^2 - z \right]^n} 
\eeq
Explicitly one has 
\beq
F_n [z] = C(w) \, z^{w+2-n} \, \frac{\Gamma (n - 2 - w)}{\Gamma (n)}  \ , \qquad n \geq 1
\label{eq:Fn}
\eeq
with $C(w) =1/(4 \pi)^{2 + w}$.
Most of the non-trivial integrals that we need to calculate have the structure of Eq.~\ref{eq:I1a}. 
In order to make progress  one needs to identify in Eq.~\ref{eq:I1a} 
the  finite part and divergent part. 
This is accomplished by using a set of recursion relations that are the subject of 
next subsection. 

\subsection{Recursion relations}

In close analogy with Ref.~\cite{Gasser:1998qt} one can define (for $m,n$ integers) :
\beq
\tilde{E}(m,n; z_\ell ) =    \int_{4}^{\infty} \frac{[ds]}{s} \, 
\int_{0}^{1} d x  \, (1 - x)^m \,   F_n [z(x,s)]     
\eeq
where $z(x,s) = z_\ell \, x^2 + s (1 - x)$. If $z_\ell \to 1$ then 
$\tilde{E}(m,n; z_\ell) \to E(m,n)$ defined in Appendix C of Ref.~\cite{Gasser:1998qt}.  
By use of integration by parts in the variable $x$ and recalling the explicit form of 
$F_n [z]$ (Eq.~\ref{eq:Fn}),  one can derive the following 
useful recursion relation:
\bea
(3 + w + m - n) \, \tilde{E}(m,n; z_\ell) &=& \frac{\Gamma (n - 2 - w)}{\Gamma (n) \Gamma (-w)} \, Q (w + 1 - n) 
\nonumber \\
&-&  n \,  z_\ell  \, \left(  \tilde{E}(m,n+1; z_\ell ) - \tilde{E}(m +2,n+1; z_\ell ) \right) 
\label{eq:recursion}
\eea
with 
\bea
Q(\alpha) &=& C(w) \, \Gamma(- w) \, \int_{4}^{\infty} \, [d s] \, s^\alpha \nonumber \\
&=&  C^2(w) \, \Gamma(-w) \, \Gamma(-1 -w - \alpha) \, \frac{\Gamma (-\alpha)}{
\Gamma (- 2 \alpha)}   ~ . 
\eea
Reassuringly, by setting $z_\ell = 1$ one recovers the result of 
Gasser-Sainio~\cite{Gasser:1998qt}. 

Eq.~\ref{eq:recursion} is useful because it allows one to express the 
integrals  $\tilde{E}(m, n \leq 2; z_\ell)$  in terms of  known divergent quantities ($Q(\alpha)$) and 
the convergent integrals $\tilde{E}(m, 3: z_\ell )$.  
Finally, let us  provide an integral representation for  $\tilde{E}(m, 3; z_\ell)$ (obtained
by setting $d=4$):  
\bea
z_\ell \, \tilde{E}(m, 3 ; z_\ell ) &=&  \frac{1}{2 (4 \pi)^4}     \,  \tilde{E}_{m} (z_\ell)   \\
\tilde{E}_m (z_\ell)  &=&  z_\ell 
\int_{4}^{\infty} \frac{ds}{s}   \left(1 - \frac{4}{s} \right)^{1/2} \, \int_0^1 \,  dx \, 
\frac{(1 -x)^m}{ z_\ell x^2 + s (1 -x)}  
\nonumber  \\ 
&=&  - 2  \, \int_0^1 \, dx  \, \frac{(1 - x)^m}{x^2} \, \left[ 
1 +  \frac{\alpha(x)}{2}  \, \log \left( 
\frac{\alpha (x) - 1}{\alpha (x) + 1}\right)
\right] 
\label{eq:Etilde}
\\
\alpha (x) &=& \left( 1 + \frac{4 (1 -x)}{z_\ell \, x^2} \right)^{1/2}
\eea
We have checked that the integrals above are indeed convergent, although we could not find 
an analytic expression for $z_\ell \neq 1$.

\subsection{Results}
We are now ready to  present results for the integrals appearing in  $T_\ell^{e^2 p^4}$.  

\bea
I_1^{(\ell)\pi \pi} &=& 
\frac{i}{(4 \pi)^4}  \left[
- \frac{m_\pi^{4w} }{(4 \pi)^{2w}}  \, \frac{ \Gamma (-w) \Gamma (-2w)  \Gamma (1-w) }{(1 + w) 
\Gamma (2 - 2 w)} +   \tilde{E}_0 (z_\ell) - \tilde{E}_2 (z_\ell)  
\right] 
\\
I_1^{(\pi)\pi \pi} &=& 
\frac{i}{(4 \pi)^4}  \left[
- \frac{m_\pi^{4w} }{(4 \pi)^{2w}}  \, \frac{ \Gamma (-w) \Gamma (-2w)  \Gamma (1-w) }{(1 + w) 
\Gamma (2 - 2 w)} -  \left(  \frac{2}{3} \pi^2  - 7  \right) \right] 
\\
I_2^{(\ell)\pi \pi} &=& i \, m_\pi^{2 + 4 w} \,  \left[   C(w) \Gamma (-w)  \right]^2 
\left[  \frac{ \Gamma(-1-w) \Gamma (-1 -2w) }{
\Gamma (-w) \Gamma (- 2 w)} 
\right. \nonumber \\
& - & \left. \frac{ 2 z_\ell}{(1+w) (2+w)} \, 
\frac{ \Gamma (-2w) \Gamma (1 -w)}{\Gamma(-w) \Gamma(2 - 2w)} \right] 
\nonumber \\
&+& i \frac{2 \,  m_\ell^2 }{(4 \pi)^4} \left[ \tilde{E}_0 (z_\ell)
-\frac{1}{2}  \tilde{E}_1  (z_\ell)
-  \tilde{E}_2 (z_\ell)
+\frac{1}{2}   \tilde{E}_3 (z_\ell)
\right]
\\
I_2^{(\pi)\pi \pi} &=& i \, m_\pi^{2 + 4 w} \,  \left[   C(w) \Gamma (-w)  \right]^2 
\left[  \frac{ \Gamma(-1-w) }{\Gamma (-w)}  + \frac{3}{2} - \frac{17}{4}\, w + \frac{59}{8} \, w^2
\right]
\\
I_3^{\pi \pi} &=& i \, m_\pi^{2 + 4 w} \, \left[   C(w) \Gamma (-w)  \right]^2  
\, \frac{\Gamma (- 1 - w)  \Gamma ( - 1 - 2 w)}{\Gamma(-w) \Gamma (-2w)} 
\\
I_4^{\pi \pi} &=& i \, m_\pi^{4 + 4 w} \, \left[   C(w) \Gamma (-w)  \right]^2  
\, \left[ \frac{\Gamma (- 1 - w)}{\Gamma(-w)} \right]^2
\\
I_5^{\pi \pi} &=&  I_4^{\pi \pi} +  i  4 \, m_\pi^{4+4w}  \Big[ z_\ell
\frac{2+w}{4 + 2w} \, \tilde{E}(0,1; z_\ell )
\nonumber \\
& - &
z_\ell^2  \left( 
\tilde{E}(0,2; z_\ell ) 
- 2 \tilde{E}(1,2 ; z_\ell )
+ \tilde{E}(2,2 ; z_\ell )
\right)
\! \! 
\Big] 
\eea
\bea
T_1^{\pi \pi} &= &
 \frac{i}{(4 \pi)^4} \frac{1/m_\pi^2}{z_\ell  - 1}  \int_{4}^{\infty} 
\frac{ds}{s} \, \left(1 - \frac{4}{s} \right)^{1/2} \, t_1^{\pi \pi} (s) 
  \equiv    \frac{i}{(4 \pi)^4 \ m_\pi^2}  \tilde{T}_1^{\pi \pi} (z_\ell) 
\\ 
t_1^{\pi \pi} (s) &=& \int_0^1 \, dx  \, \frac{1}{x}  \, \log \left(
\frac{x^2 \, z_\ell + s (1 - x)}{x^2 + s (1 - x)} 
 \right)
\nonumber \\
T_2^{\pi \pi} &=& I_1^{(\ell) \pi \pi} +  \frac{i}{(4 \pi)^4} \,  \int_{4}^{\infty} 
\frac{ds}{s} \, \left(1 - \frac{4}{s} \right)^{1/2} \, t_2^{\pi \pi} (s) 
 \  \equiv  \  I_1^{(\ell) \pi  \pi}  + \frac{i}{(4 \pi)^4}   \tilde{T}_{2}^{\pi \pi} (z_\ell) 
\\
t_2^{\pi \pi} (s) &=& 1 +  \frac{1}{ z_\ell -1} \int_0^1 \, dx  \left( 1 + \frac{s}{x} 
-\frac{s}{x^2} \right)  \, \log \left(
\frac{x^2 \, z_\ell + s (1 - x)}{x^2 + s (1 - x)} 
 \right)
\nonumber 
\eea
Note that $t_{1,2}^{\pi \pi}(s)$ can be expressed in terms of elementary functions and 
Spence functions.  The full expressions, however, are not particularly enlightening. 
Since  $t_{1,2}^{\pi \pi}(s)$ are not singular for $z_\ell \to 0$,  numerical integration 
is stable and sufficient for our purposes. 

For the two-loop integrals involving  $\bar{J}^{KK} (q^2)$  we find:
\bea
I_1^{(\ell)KK} &=& 
\frac{i}{(4 \pi)^4}  \left[
- \frac{m_K^{4w} }{(4 \pi)^{2w}}  \, \frac{ \Gamma (-w) \Gamma (-2w)  \Gamma (1-w) }{(1 + w) 
\Gamma (2 - 2 w)} +   \tilde{E}_0 (\tilde{z}_\ell)  - \tilde{E}_2  (\tilde{z}_\ell)
\right] 
\\
I_1^{(\pi)KK} &=& 
\frac{i}{(4 \pi)^4}  \left[
- \frac{m_K^{4w} }{(4 \pi)^{2w}}  \, \frac{ \Gamma (-w) \Gamma (-2w)  \Gamma (1-w) }{(1 + w) 
\Gamma (2 - 2 w)} +  
 \tilde{E}_0 (\tilde{z}_\pi)  - \tilde{E}_2  (\tilde{z}_\pi)
\right] 
\\
I_2^{(\ell)KK} &=& i \, m_K^{2 + 4 w} \,  \left[   C(w) \Gamma (-w)  \right]^2 
\left[  \frac{ \Gamma(-1-w) \Gamma (-1 -2w) }{
\Gamma (-w) \Gamma (-2 w)} 
\right. \nonumber \\
& -& \left. \frac{ 2 \tilde{z}_\ell}{(1+w) (2+w)} \, 
\frac{ \Gamma (-2w) \Gamma (1 -w)}{\Gamma(-w) \Gamma(2 - 2w)} \right] 
\nonumber \\
&+& i  \frac{2 \, m_\ell^2 }{(4 \pi)^4} \left[ \tilde{E}_0   (\tilde{z}_\ell)
-\frac{1}{2}  \tilde{E}_1 (\tilde{z}_\ell)
-  \tilde{E}_2  (\tilde{z}_\ell)
+\frac{1}{2}   \tilde{E}_3  (\tilde{z}_\ell)
\right]
\\
I_2^{(\pi)KK} &=& I_2^{(\ell)KK}  \big|_{m_\ell \to m_\pi}
\\
I_3^{KK} &=& i \, m_K^{2 + 4 w} \, \left[   C(w) \Gamma (-w)  \right]^2  
\, \frac{\Gamma (- 1 - w)  \Gamma ( - 1 - 2 w)}{\Gamma(-w) \Gamma (-2w)} 
\\
I_4^{KK} &=& i \, m_K^{4 + 4 w} \, \left[   C(w) \Gamma (-w)  \right]^2  
\, \left[ \frac{\Gamma (- 1 - w)}{\Gamma(-w)} \right]^2
\\
I_5^{KK} &=&  I_4^{KK}  \! \!  +   i  4 \, m_K^{4+4w}  \Big[ \tilde{z}_\ell 
\frac{2+w}{4 + 2w} \, \tilde{E}(0,1; \tilde{z}_\ell )
\nonumber \\
& -  &
\tilde{z}_\ell^2  \left( 
\tilde{E}(0,2 ; \tilde{z}_\ell ) 
- 2 \tilde{E}(1,2; \tilde{z}_\ell )
+ \tilde{E}(2,2 ; \tilde{z}_\ell )
\right)
\! \! 
\Big] 
\\
T_1^{KK} &=& 
 \frac{i}{(4 \pi)^4} \frac{1/m_K^2}{\tilde{z}_\ell - \tilde{z}_\pi}  \int_{4}^{\infty} 
\frac{ds}{s} \, \left(1 - \frac{4}{s} \right)^{1/2} \, t_1^{KK} (s) 
\equiv \frac{i}{(4 \pi)^4\, m_K^2} \tilde{T}_1^{KK}  (\tilde{z}_\ell, \tilde{z}_\pi)
\\ 
t_1^{KK} (s) &=& \int_0^1 \, dx  \, \frac{1}{x}  \, \log \left(
\frac{x^2 \tilde{z}_\ell + s (1 - x)}{x^2  \tilde{z}_\pi  + s (1 - x)} 
 \right)
\nonumber 
\eea
\bea
T_2^{KK} &=& I_1^{(\ell) KK} +  \frac{i}{(4 \pi)^4} \,  \int_{4}^{\infty} 
\frac{ds}{s} \, \left(1 - \frac{4}{s} \right)^{1/2} \, t_2^{KK} (s) 
 \nonumber \\
  & \equiv & \  I_1^{(\ell) KK}  + \frac{i}{(4 \pi)^4 } \,   \tilde{T}_{2}^{KK} (\tilde{z}_\ell, \tilde{z}_\pi)
\\
t_2^{KK} (s) &=& 1 +  \frac{1}{ \tilde{z}_\ell -\tilde{z}_\pi} \int_0^1 \, dx  \left( \tilde{z}_\pi
+ \frac{s}{x} 
-\frac{s}{x^2} \right)  \, \log \left(
\frac{x^2 \tilde{z}_\ell + s (1 - x)}{x^2  \tilde{z}_\pi + s (1 - x)} 
 \right)
 \nonumber 
\eea

The finite loop contributions can all be expressed in terms logarithms and combinations 
of  $\tilde{E}_n (x)$  and the following functions:
\bea
\tilde{R}_n (x) &=&  \frac{\tilde{E}_n (x)}{x}  
\label{eq:useful1}
 \\
T^{\pi \pi} (x) &=&   \tilde{T}_2^{\pi \pi} (x)  - 4 \,  \tilde{T}_1^{\pi \pi} (x) 
\label{eq:useful2}
 \\
T^{KK} (x,y) &=&   \tilde{T}_2^{KK} (x,y)  - 4 \,  \tilde{T}_1^{KK} (x,y)  ~.
\label{eq:useful}
\eea

\subsection{Standard form of two-loop integrals}
A generic two-loop contribution can be cast in the following standard form ($C(w) = 1/(4 \pi)^{2 + w}$):
\bea
I^{2-{\rm loops}} &=&  \left[ C (w) \Gamma (-w) \right]^2  \,  m^{4 w}   \  x (d)    \\
x (d) &=&  x_{0} + x_{1} \, w + x_{2 } \, w^2  + O(w^3) ~ , 
\label{eq:st2l}
\eea
with $m=m_\pi$ or  $m=m_K$.
Multiplying and dividing each contribution by   $(\mu c)^{4w}$~\cite{Gasser:1984gg}, with 
\beq
\log c =  - \frac{1}{2} \, \left( \log 4 \pi   - \gamma_E  + 1 \right) 
\eeq
and performing the  expansion around $d=4$, one finds: 
\bea
I^{2-{\rm loops}} &=& \frac{(\mu c)^{4w}}{(4 \pi)^4} \, \left[
\frac{R^{(2)}}{w^2}  +  \frac{R^{(1)}}{w}  + F    + O(w) 
\right] 
\\
R^{(2)} &=&  x_0 \\
R^{(1)} &=&  x_1 + 2 \, x_0 \, \left( \log  \frac{m^2}{\mu^2} + 1 \right)  \\
F  &=&  x_2 + 2 \, x_1 \, \left( \log  \frac{m^2}{\mu^2} + 1 \right) 
+ x_0  \left[ \frac{\pi^2}{6}   + 2 
\left( \log  \frac{m^2}{\mu^2} + 1 \right)^2  \right]
\eea

\subsection{Standard form of one-loop integrals}
The one loop diagrams with one insertion from the $O(p^4)$ effective 
lagrangian  can be cast in a useful standard form as well. 
Denoting by $L(d)$ the generic d-dimensional $p^4$ LEC, one has:
\bea
I^{1-{\rm loop}} &=&   C (w) \Gamma (-w)   \,  m^{2 w}  \ L(d)  \  y (d)    \\
y (d) &=&  y_{0} + y_{1} \, w + y_{2 } \, w^2  + O(w^3)  \\ 
L(d) &=& \frac{(\mu c)^{2w} }{(4 \pi)^2} \, \left( \frac{\Gamma}{2 w}  + 
(4 \pi)^2  L^r (\mu)    \right)  
\eea
where  $m=m_\pi$ or  $m=m_K$ and the constant $\Gamma$ determines 
the RG running of the renormalized coupling $L^r(\mu)$. 
Multiplying and dividing each contribution by   $(\mu c)^{2w}$~\cite{Gasser:1984gg}, 
and performing the  expansion around $d=4$, one finds: 
\bea
I^{1-{\rm loop}} &=& \frac{(\mu c)^{4w}}{(4 \pi)^4} \, \left[
\frac{\tilde{R}^{(2)}}{w^2}  +  \frac{\tilde{R}^{(1)}}{w}  + \tilde{F}    + O(w) 
\right] 
\\
\tilde{R}^{(2)} &= & - \frac{\Gamma \, y_0}{2}     \\
\tilde{R}^{(1)} &= & 
- (4 \pi)^2  \, L^r(\mu)   \, y_0  - \frac{ \Gamma \, (y_0 + y_1)}{2}   - 
\frac{\Gamma \, y_0}{2}    \log  \frac{m^2}{\mu^2}  \\
\tilde{F}  &=& - (4 \pi)^2  \, L^r(\mu)   \, ( y_0  + y_1)  
- \frac{ \Gamma}{24}  \left(    (6 + \pi^2) y_0   + 12 (y_1 + y_2) \right) 
\nonumber \\
&-&  \left( 2 (4 \pi)^2  L^r(\mu)   y_0 + (y_0 + y_1) \Gamma \right)  \log  \frac{m}{\mu}   
-  y_0 \Gamma \ \left(\log \frac{m}{\mu} \right)^2 
\eea
The  couplings of interest to us are $L_9$ and $L_{10}$, whose divergent parts 
are determined by:
\beq
\Gamma_{9} =  \frac{1}{4} \qquad  \Gamma_{10} = - \frac{1}{4} 
\eeq

\section{Matching calculation}
\label{sect:matchdetails}

In this Appendix we report the details of our matching calculation. 
The intermediate steps of the calculation are:
\begin{enumerate}
\item Insert the large-$N_C$ form factors of in the convolution representation of Eq.~\ref{eq:convolution}. 
\item Reduce the resulting integrals to scalar Passarino-Veltman functions~\cite{Passarino:1978jh}. 
For these we follow the convention of Kniehl~\cite{Kniehl:1993ay}.   
\item   Expand the full result  in powers of  $m_{\ell, \pi}/M_{V}$, up to order 
$(m/M_V)^2$.   This involves expanding the scalar integrals $B_0 (p^2,m_1^2,m_2^2)$ 
and $C_0( ... ) $  in powers of ratios of  the internal masses. This is trivial for  $B_0$, 
somewhat less trivial for $C_0$. 
We  derived a representation of $C_0$ as a two dimensional integral 
(see Ref.~\cite{hv1}) and used that as a starting point for the heavy mass expansion. 
\item  Subtract the ChPT$_\infty$ result from the expanded full result, thus obtaining the counterterm amplitude according to Eq.~\ref{eq:match3}. 
\end{enumerate}

\subsection{Reduction to Passarino Veltman functions}

We use the conventions of Ref.~\cite{Kniehl:1993ay} 
for the Passarino-Veltman functions,  namely:
\beq
\left\{ B_0, B_{\mu}, B_{\mu \nu} \right\} (p^2, m_1^2, m_2^2 ) =  
\int \frac{d^d q}{i  \pi^2}  \ \frac{ \left\{ 1, q_\mu  , q_\mu q_\nu  \right\} }{ [q^2 - m_1^2 + i \epsilon]
[ (q + p)^2 - m_2^2 + i \epsilon]   }
\eeq 
and 
\bea
& \left\{ C_0, C_{\mu}, C_{\mu \nu} \right\} (p^2, k^2, (p+k)^2,  m_1^2, m_2^2, m_3^2) =  &   \nonumber 
\\
& \displaystyle 
-  \int \frac{d^d q}{i  \pi^2}  \ \frac{ \left\{ 1, q_\mu  , q_\mu q_\nu  \right\} }{ [q^2 - m_1^2 + i \epsilon]
[ (q + p)^2 - m_2^2 + i \epsilon]  [ (q + p + k)^2 - m_3^2 + i \epsilon ]} ~,  & 
\eea 
with 
\bea
B_\mu  &=& p_\mu \ B_1 \\
B_{\mu \nu}   &=& p_\mu  p_\nu \ B_{21}  -   g_{\mu \nu}  \ B_{22}  \\
C_\mu  &=& p_\mu \ C_{11} + k_\mu \ C_{12} \\
C_{\mu \nu} &=&  p_\mu p_\nu \ C_{21}  + k_\mu k_\nu \ C_{22}   + (p_\mu k_\nu + k_\mu p_\nu) \ C_{23} 
- g_{\mu \nu}   C_{24}  ~. 
\eea
For the reduction of vector and tensor integrals to scalar 
Passarino-Veltman functions we have used the relations 
$(A.7), (A.8)$  and $(A.9)$ of Ref.~\cite{Kniehl:1993ay}. 

\subsubsection{$T_{V_1}$} 
In the reduction of $T_{V_1}$ we need the following tensor and vector integrals:
\bea
\int \!\!\frac{d^d q}{(2 \pi)^d}  \frac{V_1 (q^2,W^2)}{q^2 \left( q^2 - 2 q \cdot p_\ell 
\right) }  \,   q^\alpha q ^\beta   \!\!\!
&=& \!\!\! V_{\ell \ell} \,  p_\ell^\alpha  p_\ell^\beta +  V_{\nu \nu} \,  p_\nu^\alpha  p_\nu^\beta
 + V_{\nu \ell}  \, ( p_\nu^\alpha p_\ell^\beta + p_\ell^\alpha p_\nu^\beta ) 
 + V_g \,  g^{\alpha \beta} 
 \\
\int \frac{d^d q}{(2 \pi)^d} \, \frac{V_1 (q^2,W^2)}{q^2 \left( q^2 - 2 q \cdot p_\ell \right) }   \  q^\alpha & =& \!\!\!V_\pi \ p^\alpha + V_\ell \ p_\ell^\alpha 
 \eea
Using the above definitions the amplitude reads:
\bea
T_{V_1} &=&  - i e^2 T_{\ell}^{p^2} \, \Big[ 6 V_g + (m_\ell^2 - m_\pi^2) 
\Big( V_{\ell \ell} - V_{\nu \ell} \Big) \Big] 
\\
V_{\ell \ell} &=&  \frac{i}{6 (4 \pi)^2}  \left[
\frac{1}{M_V^2} \left( B_{21} (m_\ell^2, M_V^2, m_\ell^2) - B_{21} (m_\ell^2, 0 , m_\ell^2) \right)
- (1 - \kappa) \bar{C}_{21} - \kappa \bar{\bar{C}}_{21}  
\right]
\\
V_{\nu  \ell} &=&  \frac{i}{6 (4 \pi)^2}  \left[
- (1 - \kappa) \bar{C}_{23} - \kappa \bar{\bar{C}}_{23}  \right]
\\
V_{g} &=&  \frac{i}{6 (4 \pi)^2}  \left[
- \frac{1}{M_V^2} \left( B_{22} (m_\ell^2, M_V^2, m_\ell^2) - B_{22} (m_\ell^2, 0 , m_\ell^2) \right)
+ (1 - \kappa) \bar{C}_{24} +  \kappa \bar{\bar{C}}_{24}  
\right]~ ,
\eea
with 
\bea
\kappa &=& \frac{2 M_V^2 - m_\pi^2 - c_V}{M_V^2} \\
c_V &=& M_V^4 \,   \frac{N_C}{4 \pi^2 F^2}   = - 6 \, M_V^4 \, V_1 \\
\bar{C}_{ij} &=&  C_{ij} (m_\ell^2, 0, m_\pi^2, 0, m_\ell^2, M_V^2)  \\
\bar{\bar{C}}_{ij} &=&  C_{ij} (m_\ell^2, 0, m_\pi^2, M_V^2, m_\ell^2, M_V^2)  ~ .
\eea

\subsubsection{$T_{A_1}$} 
In the reduction of $T_{A_1}$ we need the following tensor, vector, and scalar  integrals:
\bea
 \int \!\! \frac{d^d q}{(2 \pi)^d}  \frac{A_1 (q^2,W^2)}{q^2 \left( q^2 - 2 q \cdot p_\ell 
\right) }   \,  q^\alpha q ^\beta \! \!\!\! 
&=& \!\!\!\! A_{\ell \ell} \,  p_\ell^\alpha  p_\ell^\beta +  A_{\nu \nu} \,  p_\nu^\alpha  p_\nu^\beta
 + A_{\nu \ell}   ( p_\nu^\alpha p_\ell^\beta + p_\ell^\alpha p_\nu^\beta ) 
 + A_g \,  g^{\alpha \beta} 
 \\
\int \frac{d^d q}{(2 \pi)^d} \, \frac{A_1 (q^2,W^2)}{q^2 \left( q^2 - 2 q \cdot p_\ell \right) }   \  q^\alpha  &=& \!\!\!\!A_\nu \ p_\nu^\alpha + A_\ell \ p_\ell^\alpha \\ 
%
%
\int \frac{d^d q}{(2 \pi)^d} \, \frac{A_1 (q^2,W^2)}{ q^2 - 2 q \cdot p_\ell  }   &=&\!\!\!\! - S_{A_1} 
\\
\int \frac{d^d q}{(2 \pi)^d} \, \frac{A_1 (q^2,W^2)}{ q^2 - 2 q \cdot p_\ell  }   \  q^\alpha 
&=& \!\!\!\! E_\nu \ p_\nu^\alpha + E_\ell \ p_\ell^\alpha 
\\
\int \frac{d^d q}{(2 \pi)^d} \, \frac{A_1 (q^2,W^2)}{ q^2 }   \  q^\alpha 
&=&  \!\!\!\! E_\pi \ p^\alpha 
\eea
Using the above definitions the amplitude reads:

\bea
T_{A_1} &=&  - i e^2 T_{\ell}^{p^2} \, \Bigg\{
S_{A_1} - E_\pi + A_\nu  (m_\pi^2 - m_\ell^2) 
\nonumber \\
&+& (d-2) 
\left[  A_g - E_\ell +  \frac{m_\pi^2 + m_\ell^2}{2} A_{\ell \ell} + 
  \frac{m_\pi^2 - m_\ell^2}{2} A_{\nu \ell}  \right]  
 \Bigg\}
\\
S_{A_1}  &=&  \frac{i}{ (4 \pi)^2}  \left[ b_1  \tilde{C}_0 + b_2 \, B_0(0,m_\ell^2, M_A^2) 
+ b_3 \, B_0 (m_\ell^2, M_V^2, m_\ell^2) \right] 
\\
A_\nu  &=&  \frac{i}{ (4 \pi)^2}  \left[
\frac{b_1}{M_V^2}   \tilde{C}_{12} -  \left(\frac{b_1}{M_V^2}  +  b_2 \right)  \, \tilde{\tilde{C}}_{12} 
 \right]
\\
A_{\ell \ell} &=&   \frac{- i}{ (4 \pi)^2}  \left[
\frac{b_3}{M_V^2} \left( B_{21} (m_\ell^2, M_V^2, m_\ell^2) - B_{21} (m_\ell^2, 0 , m_\ell^2) \right)
\right. \nonumber \\
&+& \left. 
\frac{ b_1}{M_V^2}   \tilde{C}_{21} -  \left(\frac{b_1}{M_V^2}  +  b_2 \right)   \tilde{\tilde{C}}_{21} 
\right]
\\
A_{\nu  \ell} &=&  \frac{-i}{ (4 \pi)^2}  \left[
\frac{ b_1}{M_V^2}   \tilde{C}_{23} -  \left(\frac{b_1}{M_V^2}  +  b_2 \right)   \tilde{\tilde{C}}_{23} 
\right]
\eea
\bea
A_{g} &=&  \frac{-i}{ (4 \pi)^2}  \left[
- \frac{b_3}{M_V^2} \left( B_{22} (m_\ell^2, M_V^2, m_\ell^2) - B_{22} (m_\ell^2, 0 , m_\ell^2) \right)
\right. 
\nonumber \\
&-&  \left. 
\frac{ b_1}{M_V^2}   \tilde{C}_{24} +  \left(\frac{b_1}{M_V^2}  +  b_2 \right)   \tilde{\tilde{C}}_{24} 
\right] \\
E_\ell &=&  \frac{i}{ (4 \pi)^2}  \left[ b_1 \tilde{C}_{11} - b_2 B_0(0,m_\ell^2, M_A^2) +
b_3  B_1 (m_\ell^2, M_V^2, m_\ell^2) \right]
\\
E_\pi &=&  \frac{i}{ (4 \pi)^2}  \left[ 
- \frac{ b_1}{M_V^2}  B_1 (m_\pi^2, M_V^2,M_A^2) 
 +  \left(\frac{b_1}{M_V^2}  +  b_2 \right)    B_1 (m_\pi^2, 0,M_A^2) 
\right]~, 
\eea
with
\bea
b_1 &=& M_V^2 (1 - b_2)  - M_A^2 ( 1 + b_3) \\
\tilde{C}_{ij} &=&  C_{ij} (m_\ell^2, 0, m_\pi^2, M_V^2, m_\ell^2, M_A^2)  \\
\tilde{\tilde{C}}_{ij} &=&  C_{ij} (m_\ell^2, 0, m_\pi^2, 0, m_\ell^2, M_A^2)  ~ .
\eea

\subsubsection{$T_{A_2}$} 
In the reduction of $T_{A_2}$ we need the following  vector and scalar  integrals:
\bea
\int \frac{d^d q}{(2 \pi)^d} \, \frac{A_2 (q^2,W^2)}{ q^2 - 2 q \cdot p_\ell  }   &=&  S_{A_2} 
\\
\int \frac{d^d q}{(2 \pi)^d} \, \frac{A_2 (q^2,W^2)}{ q^2 - 2 q \cdot p_\ell  }   \  q^\alpha 
&=&  F_\nu \ p_\nu^\alpha + F_\ell \ p_\ell^\alpha 
\\
\int \frac{d^d q}{(2 \pi)^d} \, \frac{A_2 (q^2,W^2)}{ q^2 }   \  q^\alpha 
&=&  \tilde{F}_\pi \ p^\alpha 
\eea
Using the above definitions the amplitude reads:
\bea
T_{A_2} &=&  i e^2 T_{\ell}^{p^2} \, \left[
- 2 S_{A_2}  + (2 - d) F_\ell  - \tilde{F}_\pi
\right]
\\
S_{A_2}  &=&  \frac{i}{ (4 \pi)^2}  \left[ (2 + d_2) M_A^2   \tilde{C}_0 
- d_2  \, B_0 (m_\ell^2, M_V^2, m_\ell^2) \right] 
\\
F_\ell &=&  - \frac{i}{ (4 \pi)^2}  \left[ (2 + d_2) M_A^2   \tilde{C}_{11} -  d_2 
 B_1 (m_\ell^2, M_V^2, m_\ell^2)  \right]
\\
\tilde{F}_\pi &=&   \frac{i}{ (4 \pi)^2}   \frac{M_A^2}{M_V^2} (2 + d_2) 
\left[   B_1 (m_\pi^2, M_V^2,M_A^2)  -   B_1 (m_\pi^2, 0,M_A^2) 
\right] ~ . 
\eea

\subsubsection{$T_{F_V}$} 
The $F_V$-induced amplitude reads:
\bea
T_{F_V} &=& 2 \frac{e^2}{(4 \pi)^2}  T_\ell^{p^2}  \Bigg\{ 
(m_\pi^2 + m_\ell^2)  \, C_0 (m_\ell^2,0,m_\pi^2, M_V^2,m_\ell^2,m_\pi^2)   
\nonumber \\
 &+& 
\frac{1}{m_\pi^2 - m_\ell^2} \left[ m_\ell^2 \,  B_0(m_\pi^2, M_V^2, m_\pi^2)  - 
m_\pi^2 \,  B_0(m_\ell^2, M_V^2, m_\ell^2)  \right] 
\Bigg\}\,.
\eea

\subsection{Expansion of the relevant  three-point scalar functions}
We use the following representation for the $C_0$ function as a basis for the large mass
expansion:
\beq
C_0 (p^2, k^2, (p+k)^2,  m_1^2, m_2^2, m_3^2)  =   \int_{0}^{1} dx \, \int_{0}^{1-x} dy \ 
\frac{1}{a x^2  + b y^2  + c  x y  + d x + e y + f} ~ , 
\eeq
with 
\bea
a &=&  (p+k)^2   \nonumber \\
b &=&   p^2   \nonumber \\
c &=&  (p+k)^2 + p^2 - k^2   \nonumber \\
d &=&  m_3^2 - m_1^2  - (p+k)^2 \nonumber \\
e &=&  m_2^2 - m_1^2  - p^2 \nonumber \\
f &=& m_1^2 \nonumber ~ . 
\eea
We then find (we give results up to  the needed order):
\bea
C_0 (m_\ell^2, 0, m_\pi^2,  M_V^2, m_\ell^2, m_\pi^2)  &= &  
\frac{1}{m_\pi^2 - m_\ell^2}  \frac{1}{M_V^2}  \left( m_\pi^2  \, \log \frac{M_V^2}{m_\pi^2} - 
m_\ell^2  \, \log \frac{M_V^2}{m_\ell^2}  \right) +  \dots
\\ 
C_0 (m_\ell^2, 0, m_\pi^2, 0, m_\ell^2, M_V^2)  &= &  
\frac{1}{M_V^2}  \left( 1 + \log \frac{M_V^2}{m_\ell^2} \right)  + 
\frac{m_\pi^2 + m_\ell^2}{4 \, M_V^4}  \left( 1 + 2  \log \frac{M_V^2}{m_\ell^2} \right) 
\nonumber \\
& + &
\frac{m_\pi^4 + m_\ell^2 m_\pi^2 +  m_\ell^4}{9 \, M_V^6}  \left( 1 + 3  \log \frac{M_V^2}{m_\ell^2} \right) 
\nonumber \\
&+& 
\frac{m_\pi^6 + m_\ell^2 m_\pi^4 + m_\ell^4 m_\pi^2 +  m_\ell^6}{16 \, M_V^8}  \left( 1 + 4  \log \frac{M_V^2}{m_\ell^2} \right)  + \dots
\\ 
C_0 (m_\ell^2, 0, m_\pi^2, M_V^2, m_\ell^2, M_V^2)  &= &  
\frac{1}{M_V^2} + \frac{1}{M_V^4} \left( \frac{5}{4} m_\ell^2  
 + \frac{1}{12} m_\pi^2  -  m_\ell^2 \log \frac{M_V^2}{m_\ell^2}   \right) 
\nonumber \\
&+&  
\frac{1}{M_V^6} \left( \frac{28}{9} m_\ell^4   
- \frac{5}{36} m_\pi^2 m_\ell^2  + \frac{1}{90} m_\pi^4  - 3  m_\ell^4 \log \frac{M_V^2}{m_\ell^2}   
\right) + \dots 
\nonumber \\
\\ 
C_0 (m_\ell^2, 0, m_\pi^2, M_V^2, m_\ell^2, M_A^2)  &= &  
\frac{1}{M_V^2} \frac{1}{z_A^2 - 1} \log z_A^2   
+ \frac{1}{M_V^4}  f^{(4)} \left( z_A^2,  m_\ell^2 ,  m_\pi^2 \right)  
\nonumber \\
&+& \! \!  \frac{1}{M_V^6}  f^{(6)} \!  \left( z_A^2,  m_\ell^2,   m_\pi^2\right)
+ \dots ~ , 
\eea
with $z_A = M_A/M_V$.  The functions $f^{(4,6)} (z_A^2, m_\ell^2, m_\pi^2)$ 
have a simple but lengthy expression: 
\bea
 f^{(4)} (z_A^2, m_\ell^2, m_\pi^2) &=& 
- \frac{m_\ell^2}{z_A^2} \log \frac{M_V^2}{m_\ell^2}     + 
\frac{\left(- 2 m_\pi^2  + m_\ell^2 (-1 + z_A^2) \right)}{2  (-1 + z_A^2)^2} 
\nonumber \\ 
&+& \frac{ \left(m_\pi^2 z_A^2 (1 + z_A^2) + m_\ell^2 (2 - 3 z_A^2  + z_A^4) \right)  }{
z_A^2 ( - 1 + z_A^2)^3}  \log z_A
\\
 f^{(6)} (z_A^2, m_\ell^2, m_\pi^2) &=& - \frac{m_\ell^4  ( 1 + 2 z_A^2)}{z_A^4}  \log \frac{M_V^2}{m_\ell^2}   +  \frac{ m_\ell^4  (3 - 3 z_A^2 + z_A^4)}{3 z_A^4 (-1 + z_A^2)^3} \log z_A^2
 \nonumber \\
 &+& \frac{  
 m_\ell^2 m_\pi^2  ( 4 - 5 z_A^2 + z_A^4)  + m_\pi^4  (1 + 4 z_A^2 + z_A^4)
 }{3  (-1 + z_A^2)^5}  \log z_A^2
\nonumber \\
&+&  \frac{ m_\ell^4  (3 - 15  z_A^2 + 10 z_A^4)}{6  z_A^2 (-1 + z_A^2)^2}
- \frac{m_\pi^4 (1 + z_A^2)}{(-1 + z_A^2)^4} 
\nonumber \\
&-& \frac{m_\ell^2 m_\pi^2 (3 + 2 z_A^2  - 7 z_A^4  + 2 z_A^6)}{6 z_A^2 ( - 1 + z_A^2)^4} ~ .
\eea

\subsection{Results}

Recalling the definition $z_A = M_A/M_V$ and neglecting as usual the $m_\ell$-independent terms 
that drop in $R_{e/\mu}$,  we find: 
\bea
T_{F_V}^{CT} &=&  T_{\ell}^{p^2} \, \frac{\alpha}{4 \pi}  \frac{m_\ell^2}{M_V^2} \, 2 \
\log \frac{M_V^2}{\mu^2}  
 \\
T_{V_1}^{CT} &=&  T_{\ell}^{p^2} \, \frac{\alpha}{4 \pi}  \frac{m_\ell^2}{M_V^2} \, 
\left[
- \frac{4}{9}   - \frac{19}{9} V_1 \, M_V^2 + \frac{4}{3} V_1 \, M_V^2 \, \log \frac{M_V^2}{\mu^2} 
\right]
 \\
T_{A_1}^{CT} &=&  T_{\ell}^{p^2} \, \frac{\alpha}{4 \pi}  \frac{m_\ell^2}{M_V^2} \, 
\Bigg[ \frac{7}{3} \left(1 - \frac{1}{z_A^2} \right)  \log \frac{M_V^2}{\mu^2} 
\nonumber \\
&-& 
\frac{ 
37 - 63 z_A^2 + 21 z_A^4 + 5  z_A^6   + 12 (7 - 10 z_A^2 + 4  z_A^4 ) \log z_A
 }{18 z_A^2 ( -1 + z_A^2)^2}
\Bigg]
\\
T_{A_2}^{CT} &=&  T_{\ell}^{p^2} \, \frac{\alpha}{4 \pi}  \frac{m_\ell^2}{M_V^2} \, 
\Bigg[ - 8 \log \frac{M_V^2}{\mu^2}  + \frac{12 - 16 z_A^2 + 4 z_A^6  
  + 4 (12 - 15 z_A^2 + 5 z_A^4) \log z_A    }{3 (-1 + z_A^2)^3}
\nonumber \\
&+&  d_2 \, \frac{8 - 14 z_A^2 + 6 z_A^4 + 2 (12 - 15 z_A^2 + 5 z_A^4) \log z_A  }{3 (-1 + z_A^2)^3}
\Bigg]~ . 
\eea
Using the input from Ref.~\cite{vcetal04} the second line of $T_{A_1}^{CT}$ should be replaced by:
\beq
\frac{ 37 - 148 z_A^2 + 168 z_A^4 - 52 z_A^6 - 5 z_A^8  + 12 (7 - 28 z_A^2 + 27 z_A^4 - 8 z_A^6) \log z_A
 }{18 z_A^2 ( -1 + z_A^2)^3}
\eeq
By comparing these expressions with the ChPT ones, one can easily verify that 
the matching procedure captures in full the single-log renormalization scale dependence, as 
expected from the $1/N_C$ expansion.

\end{document}